\pdfoutput=1
\documentclass[aps,prd,onecolumn,nofootinbib,amsmath,amssymb,floatfix,superscriptaddress,longbibliography,12pt]{revtex4-2}

\usepackage{graphicx}
\usepackage{slashed}
\usepackage{bm}
\usepackage{amsfonts}
\usepackage{mathrsfs}
\usepackage{xspace}
\usepackage{setspace}
\usepackage[colorlinks=true,linkcolor=blue,citecolor=blue,urlcolor=blue]{hyperref}

\newcommand{\gam}{\gamma}
\newcommand{\pizpiz}{\pi^{0}\pi^{0}}
\newcommand{\KsKs}{K^{0}_{S}K^{0}_{S}}
\newcommand{\KzKbz}{K^{0}\bar{K}^{0}}
\newcommand{\bk}{\mathbf{k}}

\newcommand{\kperp}{{\bk}_{\perp}}
\newcommand{\kperpi}[1]{{\bk}_{\perp #1}}
\newcommand{\mupi}{\mu_{\pi}}
\newcommand{\muK}{\mu_{K}}
\newcommand{\xbar}{\bar{x}}
\newcommand{\ybar}{\bar{y}}
\newcommand{\Hone}{\mathrm{H}_{0}^{(1)}}
\newcommand{\Jzero}{\mathrm{J}_{0}}
\newcommand{\Jone}{\mathrm{J}_{1}}
\newcommand{\Jtwo}{\mathrm{J}_{2}}
\newcommand{\Kzero}{\mathrm{K}_{0}}
\newcommand{\TildeT}{\widetilde{T}}
\newcommand{\LamQCD}{\Lambda_{\mathrm{QCD}}}

\begin{document}

\title{Twist-3 contributions to
       $\gam\gam\to\pi^{0}\pi^{0},\,K_{S}^{0}K_{S}^{0}$
       in $k_{T}$ factorization}

\author{Jun-Kang He}
\email{hejk@hbnu.edu.cn}
\affiliation{College of Physics and Electronic Science, Hubei Normal University,
             Huangshi 435002, China}

\author{Cong Wang}
\email{wangj@hbnu.edu.cn}
\affiliation{College of Physics and Electronic Science, Hubei Normal University,
             Huangshi 435002, China}

\begin{abstract}
We compute the cross sections for the two-photon processes
$\gam\gam\to\pi^{0}\pi^{0}$ and $\gam\gam\to K_{S}^{0}K_{S}^{0}$ in
$k_{T}$ factorization, including the chirally enhanced two-parton twist-3
light cone distribution amplitudes.  For these charge-suppressed neutral
channels the twist-3 cross sections exceed the twist-2 ones by close to an
order of magnitude in the intermediate-energy region, bringing the
predictions much closer to the Belle data,
the residual underestimate being plausibly attributable to higher-order
QCD corrections.  The calculation reproduces the measured angular
distributions and the energy dependence of the charged channels and the
neutral pion, though not the steeper fall of the neutral kaon.  The
neutral-to-charged ratios are the most discriminating observables.  They
depend strongly on energy in the data, whereas our calculation, like other
approaches in the literature, yields a nearly flat ratio.  Finally, in a
phenomenological discussion, we combine our contribution with the soft
handbag contribution and largely reproduce the observed energy dependence,
suggesting that the hard and soft contributions are comparably important in
the few-GeV region.
\end{abstract}

\maketitle

\section{Introduction}
\label{sec:intro}

The exclusive two-photon annihilation into meson pairs at large invariant
mass, $\gam\gam\to M\overline{M}$, has long served as one of the cleanest
testing grounds for the application of perturbative quantum
chromodynamics (pQCD) to exclusive hard processes.  Because the initial
state is purely electromagnetic, all nonperturbative dynamics resides
in the meson light cone distribution amplitudes (LCDAs) of the final
state, so the process probes the LCDAs essentially directly---without
the additional nonperturbative complications---parton distribution
functions, soft spectators, and initial-state radiation---that accompany
an initial-state hadron, as in deep inelastic or hadron-hadron
scattering.  In the asymptotic regime
$s,|t|,|u|\gg\LamQCD^{2}$, Brodsky and
Lepage~\cite{Brodsky:1981rp,Lepage:1980fj} demonstrated that the
amplitude factorizes into a perturbatively computable hard kernel for
the partonic process $\gam\gam\to q\bar q q\bar q$ convoluted with two
single-meson twist-2 LCDAs, yielding the very sharp leading-twist (LT)
prediction $d\sigma/d|\cos\theta^{*}|\sim\sin^{-4}\theta^{*}/Q^{6}$ for
the charged pairs and a total cross section falling as
$\sigma(Q)\sim Q^{-6}$.  The neutral channels are suppressed for a
reason common to both.  A neutral pseudoscalar meson is built from
valence (anti)quarks of equal charge, so its charge $e_{1}-e_{2}$
vanishes.  Because the leading $\sin^{-4}\theta^{*}$ term of the Brodsky-Lepage
amplitude scales as $(e_{1}-e_{2})^{4}$, it cancels identically for both
neutral channels, which are then governed by the subleading
$g(\theta^{*})$ pieces, so that the rate is strongly suppressed and the angular
distribution departs from the simple $\sin^{-4}\theta^{*}$
law~\cite{Brodsky:1981rp,Chernyak:2006dk,Chen:2006gy}.  Detailed LT
calculations give a small ratio
$\sigma(\pi^{0}\pi^{0})/\sigma(\pi^{+}\pi^{-})\simeq 0.04$--$0.07$~\cite{Chernyak:2006dk,Chernyak:2014wra,Uehara:2009cka}.
Over the past four decades these
channels have been measured with steadily improving precision, from the
early TPC/Two-Gamma~\cite{Aihara:1986qk} and ALEPH~\cite{Heister:2003ae}
experiments to the very high-statistics Belle
program~\cite{Nakazawa:2004gu,Mori:2007bu,Chen:2006gy,Uehara:2008ep,Uehara:2009cka,Uehara:2009cf,Uehara:2010mq,Uehara:2013mbo,Masuda:2015yoh,Masuda:2017xey},
which now spans the entire relevant range $0.6\lesssim Q\lesssim
4.1~\mathrm{GeV}$ for the charged channels $\pi^{+}\pi^{-},K^{+}K^{-}$
and for the neutral channels $\pi^{0}\pi^{0}$, $K_{S}^{0}K_{S}^{0}$,
$\eta\pi^{0}$ and $\eta\eta$.  Confronting these data with the pure LT
predictions in the perturbative window $Q\sim 2$--$4~\mathrm{GeV}$
exposes a striking and systematic discrepancy.  The
measured cross sections exceed the LT results by roughly an order of
magnitude for the charged
pairs~\cite{Nakazawa:2004gu,Brodsky:1981rp,Coriano:1998ge,Hsieh:2004ee,Wang:2015mod}
and by even more for the charge-suppressed neutral pairs.  The measured
ratios sharpen the contrast: $\sigma(\pi^{0}\pi^{0})/\sigma(\pi^{+}\pi^{-})$
stabilizes around $0.3$--$0.5$~\cite{Uehara:2008ep,Uehara:2009cka,Nakazawa:2004gu},
far above the LT expectation of $0.04$--$0.07$, while
$\sigma(K_{S}^{0}K_{S}^{0})/\sigma(K^{+}K^{-})$ falls
steeply from $\sim 0.13$ to $\sim 0.01$ across the same energy
interval~\cite{Chen:2006gy}.  Since the $K^{0}$ is charge-suppressed for
exactly the same reason as the $\pi^{0}$, the LT mechanism predicts a
comparably small and nearly energy-independent neutral-kaon ratio, so that
neither its magnitude at low $Q$ nor its pronounced energy variation is
reproduced.  The energy power-laws extracted from the
data, $\sigma(Q)\sim Q^{-n}$ with $n_{\pi^{0}\pi^{0}}\simeq 7$--$8$ and
$n_{K_{S}^{0}K_{S}^{0}}\simeq 10$--$11$~\cite{Uehara:2009cka,Uehara:2013mbo},
the latter in particular being far steeper than the canonical $n=6$.  Taken
together, the substantial underestimate of the cross sections across
all channels indicates that the leading-twist amplitude does not capture
the dominant contribution to these processes in the present energy
window, and that subleading power corrections play a quantitatively
decisive role rather than a perturbative one.

Several theoretical strategies have been put forward to bridge this
gap.  At the level of pure collinear pQCD, the next-to-leading-order
(NLO) calculation of Duplan\v{c}i\'c and Ni\v{z}i\'c~\cite{Duplancic:2006nv}
yields sizable $K$-factors.  At the renormalization scale $\mu_{R}^{2}=Q^{2}$
the NLO term amounts to about $90\%$ of the leading order at
$Q=4~\mathrm{GeV}$ (a $K$-factor of order $1.9$), so that the
perturbative series converges slowly across the Belle window.  The NLO
predictions nonetheless still lie below the measured cross
sections, which suggests that in this regime important contributions arise
from still higher orders---a point we
return to in connection with our own results.  Within the soft handbag mechanism developed by Diehl, Kroll, and
collaborators~\cite{Diehl:2001fv,Diehl:2009yi,Kroll:2012iv}, the hard
subprocess $\gam\gam\to q\bar q$ is folded with a phenomenological
$q\bar q\to M\overline{M}$ form factor.  This approach reproduces the charged
cross sections through phenomenological
nonperturbative inputs fitted to the data, and predicts nearly
energy-independent SU(3) ratios---$\sigma(\pi^{0}\pi^{0})/\sigma(\pi^{+}\pi^{-})\sim
1/2$, in fair agreement with the pion data, and
$\sigma(K_{S}^{0}K_{S}^{0})/\sigma(K^{+}K^{-})\approx 2/25$ in the SU(3)
limit~\cite{Diehl:2001fv,Diehl:2009yi}.  The latter is in tension with
the pronounced fall of the measured ratio from $\sim
0.13$ to $\sim 0.01$~\cite{Chen:2006gy}, which is not easily
accommodated within the handbag picture even once SU(3)-breaking is
included, and which suggests that further contributions are at
play~\cite{Chen:2006gy,Chernyak:2014wra}.  Below the perturbative window the
amplitudes are dominated by hadronic resonances and have been studied
in detail using dispersive and Roy-Steiner
techniques~\cite{Hoferichter:2011wk,GarciaMartin:2010cw,Dai:2014lza,Danilkin:2018qfn,Surovtsev:2021kud}
and in coupled-channel chiral models~\cite{Klevansky:2016abt}.

To compute the subleading contributions systematically we adopt the
$k_{T}$-factorization (Li-Sterman) framework, originally developed for
the pion electromagnetic form
factor~\cite{Botts:1989kf,Li:1992nu,Jakob:1993iw} and applied to
two-photon meson-pair production in
Refs.~\cite{Coriano:1998ge,Hsieh:2004ee}.  Supplemented with the chirally
enhanced two-parton twist-3 LCDAs, it was applied in
Refs.~\cite{Wang:2015mod,Wang:2019trq} to the Belle
charged-channel data $\gam\gam\to\pi^{+}\pi^{-},K^{+}K^{-}$, and we extend
it here to the neutral channels.  Retaining the parton
transverse momentum $\kperp$, together with the
Sudakov~\cite{Botts:1989kf,Li:1992nu} and
threshold~\cite{Li:1998is,Kurimoto:2001zj,Li:2001ay} resummations,
regulates the twist-3 end point region and makes the framework well
suited to the intermediate-energy window $Q\sim 2$--$4~\mathrm{GeV}$,
where the smallest partonic virtualities approach $\LamQCD^{2}$ and a
purely collinear treatment is no longer reliable.  The twist-3 terms are
numerically decisive in this window because they are weighted by the
chiral mass $\mu_{M}=m_{M}^{2}/(m_{q_{1}}+m_{q_{2}})$, $M\in\{\pi,K\}$,
which by the Gell-Mann-Oakes-Renner (GMOR)
relation~\cite{GellMann:1968rz,Leutwyler:1996qg} is numerically as large
as $\mu_{M}(2~\mathrm{GeV})\simeq 2.6~\mathrm{GeV}$.  The
two-parton twist-3 contributions then enter relative to the
leading-twist ones with the chiral-enhancement parameter
$4\mu_{M}^{2}/Q^{2}$, which is of order unity across the
window~\cite{Ball:1998je,Ball:2006wn,Beneke:2000wa}.

For the neutral channels the transverse-momentum and twist-3 effects do
not merely compete with the leading twist but carry the cross section,
since the charge suppression noted above removes the would-be dominant
leading-twist amplitude.  Ratios of the neutral to the charged cross
sections are theoretically more robust than absolute rates, since the
overall normalization and many input and systematic
uncertainties largely cancel.  It is therefore significant that the
measured neutral-to-charged ratios vary so rapidly with energy.
Specifically, the pion ratio turns upward at the upper end of the window,
and the kaon ratio falls by an order of magnitude across it.  Such a strong energy
dependence of a quantity in which the leading effects largely cancel is a
strong indication that the subleading contributions of higher twist,
intrinsic transverse momentum, and higher-order QCD corrections are
important in this regime.  Their roles are distinct.  Throughout this
window the chiral enhancement renders the twist-3 contributions
dominant, exceeding the leading-twist ones and adding constructively to
them, while the transverse-momentum and Sudakov effects, which by
themselves already modify the twist-2 cross sections
substantially~\cite{Wang:2019trq}, act in the twist-3 sector as the
joint regulator of the end point region.  The higher-order QCD
corrections are sizable in collinear factorization, as noted
above, whereas for the pion electromagnetic form
factors in $k_{T}$ factorization the partial cancellation between the
twist-2 and twist-3 NLO corrections leaves only a moderate net
effect~\cite{Cheng:2014gba,Cheng:2015qra}.  Even so, the strong and
channel-dependent energy dependence of the ratios may signal the
coexistence of several competing mechanisms, perturbative hard
scattering alongside soft end point or handbag dynamics, in the
intermediate-energy region.

The present paper addresses the natural next step, a complete
$k_{T}$-factorization calculation, including the chirally enhanced
two-parton twist-3 LCDAs, of the neutral channels
$\gam\gam\to\pi^{0}\pi^{0}$ and $\gam\gam\to K_{S}^{0}K_{S}^{0}$.  These
neutral channels are the most stringent discriminators among the
competing dynamical mechanisms, whose predictions for the ratio
$\sigma(\pizpiz)/\sigma(\pi^{+}\pi^{-})$ quoted above differ by an order
of magnitude, while the measured values lie in between.  A
first-principles perturbative calculation with consistent power
corrections is therefore necessary to disentangle the perturbative
from the soft content in this regime.  Moreover, in contrast to the
charged channels the neutral amplitudes carry a distinctive
charge-flavor structure---the flavor superposition
$\pi^{0}=(u\bar u-d\bar d)/\sqrt{2}$ and the $CP$ projection of the
$K_{S}^{0}K_{S}^{0}$ final state---which modifies the color-flavor
combinatorics and the symmetry pattern of the four diagrammatic groups
in nontrivial ways and propagates into the helicity hard kernels, so
that the neutral channels probe the twist-3 dynamics from an angle
complementary to that of the charged channels studied in
Ref.~\cite{Wang:2019trq}.

This paper is organized as follows.  Section~\ref{sec:formalism} sets up
the theoretical framework, from the kinematics and the LCDAs of $\pi^{0}$,
$K^{0}$ and $\bar K^{0}$ to the $k_{T}$-factorization helicity amplitudes
and the twist-3 hard kernels of the four diagrammatic groups, and the
extension to $\gam\gam\to\KsKs$.  Section~\ref{sec:numerics} presents the
numerical analysis.  It first defines the cross sections and collects the
nonperturbative inputs, then gives the twist-2 and twist-3 predictions
for the cross sections, their energy dependence, the normalized
angular distributions and the neutral-to-charged ratios in comparison
with the Belle data, and closes with a discussion in which a soft handbag
contribution is added coherently to the hard amplitude to account for the
observed energy dependence of the ratios.  Section~\ref{sec:summary}
gives a summary.

\section{Theoretical framework}
\label{sec:formalism}

\subsection{Kinematics and conventions}
\label{subsec:kinematics}

We work in the center-of-mass (CM) frame of the two-photon system,
with the outgoing neutral meson $M^{0}(p_{3})$ directed along the positive
$z$-axis.  We employ light cone coordinates
$A^{\mu}=(A^{+},A^{-},\mathbf{A}_{\perp})$ with
$A^{\pm}\equiv A^{0}\pm A^{3}$.  In these coordinates the four external
momenta of the process
$\gam(p_{1},\lambda_{1})\,\gam(p_{2},\lambda_{2})
\to M^{0}(p_{3})\,\overline{M}^{0}(p_{4})$ are
\begin{equation}
\label{eq:kinmomenta}
\begin{aligned}
p_{1}^{\mu}&=(2\omega c^{2},\,2\omega s^{2},\,\mathbf{p}_{\perp}),
&\quad
p_{2}^{\mu}&=(2\omega s^{2},\,2\omega c^{2},\,-\mathbf{p}_{\perp}),\\
p_{3}^{\mu}&=(2\omega,\,0,\,\mathbf{0}_{\perp}),
&\quad
p_{4}^{\mu}&=(0,\,2\omega,\,\mathbf{0}_{\perp}),
\end{aligned}
\end{equation}
with the half-CM energy
$\omega=Q/2$,
$s\equiv\sin(\theta^{*}/2)$, $c\equiv\cos(\theta^{*}/2)$, and the transverse
photon momentum $\mathbf{p}_{\perp}=(2\omega s c,0)$.  Here $\theta^{*}$ is
the polar scattering angle of the meson $M^{0}(p_{3})$ with respect to the
incoming photon $\gam(p_{1})$, and we neglect the meson masses
($m_{M}^{2}\ll Q^{2}$) in the external kinematics.  The corresponding
photon polarization vectors are
\begin{equation}
\label{eq:polarization}
\begin{aligned}
\varepsilon^{\mu}(p_{1},+)&=\tfrac{1}{\sqrt{2}}(2sc,-2sc,s^{2}-c^{2},-i),\\
\varepsilon^{\mu}(p_{1},-)&=\tfrac{1}{\sqrt{2}}(-2sc,2sc,c^{2}-s^{2},-i),\\
\varepsilon^{\mu}(p_{2},+)&=\tfrac{1}{\sqrt{2}}(-2sc,2sc,c^{2}-s^{2},-i),\\
\varepsilon^{\mu}(p_{2},-)&=\tfrac{1}{\sqrt{2}}(2sc,-2sc,s^{2}-c^{2},-i).
\end{aligned}
\end{equation}

For each outgoing meson, the constituent quark and antiquark are
labeled by their longitudinal momentum fractions $x$, $\xbar=1-x$ in
$M^{0}(p_{3})$ and $y$, $\ybar=1-y$ in $\overline{M}^{0}(p_{4})$, together with intrinsic
transverse momenta $\kperpi{1}$ and $\kperpi{2}$
\begin{equation}
\label{eq:partonmomenta}
\begin{aligned}
k_{1}^{\mu}&=(2x\omega,\,0,\,\kperpi{1}),
&\quad
k_{2}^{\mu}&=(2\xbar\omega,\,0,\,-\kperpi{1}),\\
k_{3}^{\mu}&=(0,\,2y\omega,\,\kperpi{2}),
&\quad
k_{4}^{\mu}&=(0,\,2\ybar\omega,\,-\kperpi{2}).
\end{aligned}
\end{equation}
The Mandelstam invariants reduce in this parametrization to
$t\!=\!-Q^{2}s^{2}$ and $u\!=\!-Q^{2}c^{2}$, the squared center-of-mass
energy being $Q^{2}$ (so that $t+u=-Q^{2}$).  All ratios of hard scales we shall encounter are
functions of $(x,y,\theta^{*})$ alone, which substantially organizes the
analytic structure of the hard kernels.

The reliability of the perturbative treatment is controlled not by
$Q^{2}$ alone but by the smaller momentum transfers $|t|$ and $|u|$,
which set the virtualities of the internal quark and gluon
propagators.  Since $\min(|t|,|u|)=Q^{2}\min(s^{2},c^{2})$ becomes
parametrically small near the forward and backward directions, the
hard-scattering picture inevitably degrades at large scattering
angles, where (i) the smallest propagator virtuality drops toward
$\LamQCD^{2}$ and the running coupling is probed in the
nonperturbative regime, (ii) the soft end point regions of the parton
momentum fractions become enhanced, and (iii) the soft handbag/overlap
mechanism competes with the leading hard-scattering
contribution~\cite{Diehl:2001fv,Hsieh:2004ee}.  To stay within the
domain where the perturbative calculation is trustworthy---and to
match the kinematic range over which the Belle Collaboration extracts
the cross sections for theory comparison---we restrict the
scattering angle throughout this work to
$|\cos\theta^{*}|<0.6$~\cite{Uehara:2009cka,Uehara:2013mbo}, for which
$\min(|t|,|u|)>0.2\,Q^{2}$ remains safely above the soft scale across
the Belle energy window.

\begin{figure}[!htb]
  \centering
  \includegraphics[width=0.75\textwidth]{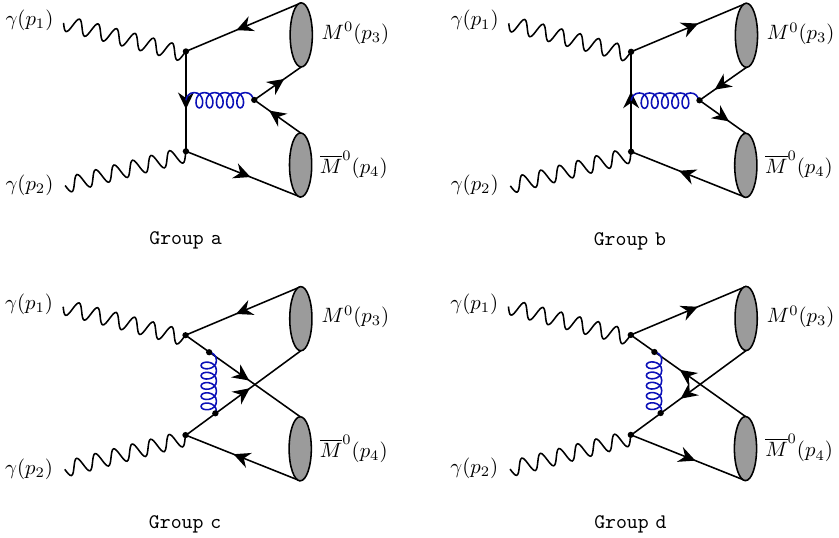}
  \caption{The four basic Feynman diagrams contributing to the
    partonic amplitude $\gam(p_{1})\,\gam(p_{2})\to
    q(k_{1})\bar q(k_{2})\,q(k_{3})\bar q(k_{4})$ for
    $\gam\gam\to M^{0}(p_{3})\overline{M}^{0}(p_{4})$.  The total numbers of diagrams in
    groups $a$, $b$, $c$ and $d$ are $6,6,4,4$, respectively, and the
    grouping is dictated by the virtuality of the internal gluon
    propagator.}
  \label{fig:feyn}
\end{figure}

At the parton level the amplitude is built from the lowest-order
subprocess
$\gam(p_{1})\,\gam(p_{2})\to q(k_{1})\bar q(k_{2})\,q(k_{3})\bar q(k_{4})$,
in which the two photons couple to the quark lines and a single gluon is
exchanged to bind the two $q\bar q$ pairs into the outgoing mesons.  The
contributing diagrams fall into four topological groups $a,b,c,d$ shown in
Fig.~\ref{fig:feyn}, distinguished by the virtuality of the internal gluon:
the gluon is attached to a single quark line in groups $a$ and $b$ (the
direct topology) and is exchanged between the two quark lines in groups $c$
and $d$ (the cross-quark topology).  These topologies set the analytic
structure of the hard kernels evaluated in
Sec.~\ref{subsec:hardkernels}.

\subsection{Light cone distribution amplitudes}
\label{subsec:lcda}

The nonperturbative content of each outgoing meson is encoded in its
light cone distribution amplitudes (LCDAs), defined by gauge-invariant
nonlocal bilinear quark-antiquark matrix elements between the meson
state and the vacuum.  At twist-3 accuracy three two-parton LCDAs enter---one twist-2 (axial-vector projection) and two twist-3
(pseudoscalar and pseudotensor projections).  The three-parton twist-3
LCDA is suppressed by $\eta_{3M}=f_{3M}/(f_{M}\mu_{M})\sim 10^{-2}$ and is neglected
throughout this work~\cite{Ball:1998je,Ball:2006wn,Beneke:2000wa,Wang:2019trq}.
The convergence of the twist expansion beyond this accuracy has been
examined for the closely related pion and kaon electromagnetic form
factors, where recent $k_{T}$-factorization analyses including the
higher-twist contributions from both the two-particle and three-particle
configurations find a satisfactory convergence beyond
twist~3~\cite{Cheng:2019ruz,Chai:2024tss,Chai:2025xuz}, the
three-particle twist-3 terms being suppressed by $\eta_{3M}$ and the
twist-4 corrections carrying a further $1/Q^{2}$ power suppression
without chiral enhancement.  Given the similarity between those
convolutions and the present helicity amplitudes, the two-parton twist-2
and twist-3 contributions retained here are expected to capture the
dominant part of the twist expansion in the intermediate-energy region.

By isospin symmetry the neutral and charged pions share the same decay
constant, $f_{\pi^{0}}=f_{\pi^{\pm}}\equiv f_{\pi}$, and the same set
of LCDAs, and we therefore define the $\pi^{0}$ LCDAs through the
isospin-symmetric matrix elements normalized directly to the physical
decay constant $f_{\pi}$.  With $z\equiv z_{1}-z_{2}$ a lightlike
separation, the relevant matrix elements
read~\cite{Braun:1988qv,Braun:1989iv,Ball:1998je,Beneke:2000wa,Wei:2002iu}
\begin{align}
\label{eq:pionmatrix}
\langle\pi^{0}(p)|\bar q(z_{1})\gamma_{\mu}\gamma_{5}q(z_{2})|0\rangle
   &=-if_{\pi}\,p_{\mu}\!\int_{0}^{1}\!\!\mathrm{d}x\,
       e^{i(xp\cdot z_{1}+\xbar p\cdot z_{2})}
       \phi_{\pi}^{\,\pi}(x,\mu),\nonumber\\[2pt]
\langle\pi^{0}(p)|\bar q(z_{1})i\gamma_{5}q(z_{2})|0\rangle
   &=f_{\pi}\,\mupi\!\int_{0}^{1}\!\!\mathrm{d}x\,
       e^{i(xp\cdot z_{1}+\xbar p\cdot z_{2})}
       \phi_{\pi}^{p}(x,\mu),\nonumber\\[2pt]
\langle\pi^{0}(p)|\bar q(z_{1})i\gamma_{5}\sigma_{\mu\nu}q(z_{2})|0\rangle
   &=-f_{\pi}\,\mupi\bigl(p_{\mu}z_{\nu}-p_{\nu}z_{\mu}\bigr)
       \!\int_{0}^{1}\!\!\mathrm{d}x\,e^{i(xp\cdot z_{1}+\xbar p\cdot z_{2})}
       \frac{\phi_{\pi}^{\sigma}(x,\mu)}{6},
\end{align}
where $f_{\pi}=130.2(8)~\mathrm{MeV}$~\cite{ParticleDataGroup:2024cfk}
and $\bar q\,\Gamma\,q=(\bar u\,\Gamma\,u-\bar d\,\Gamma\,d)/\sqrt{2}$ denotes the isospin-projected light-quark
bilinear appropriate to the $\pi^{0}$
state.  The superscript on each LCDA labels the two-parton amplitude by the
Dirac structure of its defining matrix element rather than by a meson:
$\phi_{M}^{\,\pi}$ is the leading twist-2 (axial-vector) LCDA, while
$\phi_{M}^{p}$ and $\phi_{M}^{\sigma}$ are the twist-3 pseudoscalar and
pseudotensor LCDAs.  The same convention is used for the kaon below, so
that, for instance, $\phi_{K}^{\,\pi}$ denotes the twist-2 kaon LCDA.  The resolution of this isospin structure into the individual
$u$- and $d$-quark contributions, together with the photon electric
charges, is carried out at the level of the hard-scattering amplitude
in Sec.~\ref{subsec:hardkernels}.  The chirally enhanced mass parameter
\begin{equation}
\label{eq:mupiGMOR}
\mupi(\mu)=\frac{m_{\pi}^{2}}{m_{u}(\mu)+m_{d}(\mu)}
=-\frac{2\langle\bar qq\rangle(\mu)}{f_{\pi}^{2}}+\mathcal{O}(m_{q})
\end{equation}
is, by the isospin symmetry of the matrix elements
\eqref{eq:pionmatrix}, common to both flavor components of the
$\pi^{0}$, and in the isospin-symmetric limit it is fixed by the
GMOR relation
$f_{\pi}^{2}m_{\pi}^{2}=-2(m_{u}+m_{d})\langle\bar
qq\rangle$~\cite{GellMann:1968rz,Leutwyler:1996qg}, where
$\langle\bar qq\rangle$ is the single-flavor quark condensate.  Although
$\mupi$ multiplies a formally power-suppressed twist-3 contribution,
the smallness of the current-quark mass $(m_{u}+m_{d})$ in its
denominator makes it numerically large,
$\mupi\simeq 2.6~\mathrm{GeV}$ at $\mu=2~\mathrm{GeV}$, comparable to the
hard scale $Q\sim 2$--$4~\mathrm{GeV}$ of the Belle window.  This chiral
enhancement, introduced in Sec.~\ref{sec:intro}, is why the two-parton
twist-3 sector contributes significantly in the intermediate-energy
region.

The shape of each LCDA is organized by the conformal expansion.  At the
next-to-leading conformal
spin~\cite{Braun:1988qv,Braun:1989iv,Ball:1998je,Ball:2006wn}, the
twist-2 LCDA of a pseudoscalar meson $M$ is
\begin{equation}
\label{eq:phi2Gegenbauer}
\phi^{\,\pi}_{M}(x,\mu)=6x\xbar\,\Bigl[1
   +\sum_{n=1}^{\infty}a_{n}^{M}(\mu)\,C_{n}^{3/2}(2x-1)\Bigr],
\end{equation}
with $C_{n}^{3/2}$ the Gegenbauer polynomials.  For the
isospin-symmetric pion, $G$-parity enforces $a_{2n+1}^{\pi}=0$, and we
retain $a_{2}^{\pi}$ and $a_{4}^{\pi}$.  The two-parton twist-3 LCDAs of
the pion, including the meson-mass corrections of
Refs.~\cite{Ball:1998je,Ball:2006wn}, are
\begin{align}
\label{eq:phipPion}
\phi_{\pi}^{p}(x,\mu)&=1+\bigl[30\,\eta_{3\pi}(\mu)
   -\tfrac{5}{2}\rho_{\pi}^{2}(\mu)\bigr]C_{2}^{1/2}(2x-1)\\
&\quad-\bigl[3\,\eta_{3\pi}(\mu)\omega_{3\pi}(\mu)
   +\tfrac{27}{20}\rho_{\pi}^{2}(\mu)
   +\tfrac{81}{10}\rho_{\pi}^{2}(\mu)a_{2}^{\pi}(\mu)\bigr]
   C_{4}^{1/2}(2x-1),\nonumber\\
\label{eq:phisigPion}
\phi_{\pi}^{\sigma}(x,\mu)&=6x\xbar
   \Bigl\{1+\bigl[5\eta_{3\pi}(\mu)
   -\tfrac{1}{2}\eta_{3\pi}(\mu)\omega_{3\pi}(\mu)
   -\tfrac{7}{20}\rho_{\pi}^{2}(\mu)
   -\tfrac{3}{5}\rho_{\pi}^{2}(\mu)a_{2}^{\pi}(\mu)\bigr]
   C_{2}^{3/2}(2x-1)\Bigr\},
\end{align}
where the Legendre polynomials are
$C_{2}^{1/2}(\xi)=\tfrac{1}{2}(3\xi^{2}-1)$ and
$C_{4}^{1/2}(\xi)=\tfrac{1}{8}(35\xi^{4}-30\xi^{2}+3)$, the three-particle
coupling is $\eta_{3\pi}=f_{3\pi}/(f_{\pi}\mupi)$, $\omega_{3\pi}$ is
its gluon-momentum-fraction parameter, and
$\rho_{\pi}^{2}=(m_{u}+m_{d})^{2}/m_{\pi}^{2}$ is the meson-mass
correction parameter, which is numerically tiny but kept for
consistency because at twist-3 it formally enters at the same order as
$f_{3\pi}$~\cite{Ball:1998je}.

The neutral-kaon LCDAs are defined analogously.  Adopting the standard
convention in which the strange quark sits in the $q_{2}$
position~\cite{Ball:2006wn}, the LCDAs of $\bar K^{0}=s\bar d$ coincide,
in the exact-isospin limit, with the conventionally defined LCDAs of
the $K^{-}=s\bar u$ meson.  The relevant matrix elements
read~\cite{Ball:2006wn,Beneke:2000wa}
\begin{align}
\label{eq:K0matrix}
\langle\bar K^{0}(p)|\bar s(z_{1})\gamma_{\mu}\gamma_{5}d(z_{2})|0\rangle
   &=-if_{K}\,p_{\mu}\!\int_{0}^{1}\!\!\mathrm{d}x\,
       e^{i(xp\cdot z_{1}+\xbar p\cdot z_{2})}
       \phi_{K}^{\,\pi}(x,\mu),\nonumber\\
\langle\bar K^{0}(p)|\bar s(z_{1})i\gamma_{5}d(z_{2})|0\rangle
   &=f_{K}\muK\!\int_{0}^{1}\!\!\mathrm{d}x\,
       e^{i(xp\cdot z_{1}+\xbar p\cdot z_{2})}
       \phi_{K}^{p}(x,\mu),\nonumber\\
\langle\bar K^{0}(p)|\bar s(z_{1})i\gamma_{5}\sigma_{\mu\nu}d(z_{2})|0\rangle
   &=-f_{K}\muK\bigl(p_{\mu}z_{\nu}-p_{\nu}z_{\mu}\bigr)
       \!\int_{0}^{1}\!\!\mathrm{d}x\,e^{i(xp\cdot z_{1}+\xbar p\cdot z_{2})}
       \frac{\phi_{K}^{\sigma}(x,\mu)}{6},
\end{align}
with the kaon decay constant
$f_{K}=155.7(3)~\mathrm{MeV}$~\cite{ParticleDataGroup:2024cfk}
and the chirally enhanced mass parameter
$\muK=m_{K^0}^{2}/(m_{d}+m_{s})$.  The LCDAs of the
strangeness-conjugate state $K^{0}=d\bar s$ follow from charge
conjugation, which acts on the longitudinal momentum fraction as
$x\leftrightarrow\xbar$,
\begin{equation}
\label{eq:K0conj}
\phi_{K^{0}}^{\,i}(x,\mu)=\phi_{\bar K^{0}}^{\,i}(\xbar,\mu),
\qquad i\in\{\pi,p,\sigma\},
\end{equation}
with the odd Gegenbauer moments and the $SU(3)$-breaking twist-3
parameters changing sign accordingly~\cite{Ball:2006wn}.  The
relation~\eqref{eq:K0conj} is the origin of the argument $(1-x)$
carried by the $K^{0}$ leg in the helicity
amplitude~\eqref{eq:HelicityAmpKK} below.  In the twist-2 conformal
expansion~\eqref{eq:phi2Gegenbauer} the kaon retains, besides
$a_{2}^{K}$, a nonzero $a_{1}^{K}$ generated by $SU(3)$-flavor breaking.

The kaon twist-3 LCDAs are richer in structure because
$SU(3)$-flavor breaking lifts the $G$-parity selection rules that
forbid odd-conformal-spin terms in the pion case.  Two new
dimensionless quark-mass parameters,
\begin{equation}
\label{eq:rhoKpm}
\rho_{K}^{+}=\frac{(m_{s}+m_{q})^{2}}{m_{K}^{2}},
\qquad
\rho_{K}^{-}=\frac{m_{s}^{2}-m_{q}^{2}}{m_{K}^{2}},
\qquad m_{q}=\tfrac{1}{2}(m_{u}+m_{d}),
\end{equation}
encode the symmetric and antisymmetric (G-parity violating) parts of
the quark-mass-induced corrections.  In the Ball-Braun-Lenz (BBL)
parametrization~\cite{Ball:2006wn} the two-parton twist-3 LCDAs of
$\bar K^{0}$ at next-to-leading conformal spin
read~\cite[Eqs.(3.25)--(3.26)]{Ball:2006wn}
\begin{align}
\label{eq:phipKaon}
\phi_{K}^{p}(x,\mu)
&=1+3\rho_{K}^{+}\bigl(1+6\,a_{2}^{K}-9\,\rho_{K}^{-}a_{1}^{K}\bigr)
+\Bigl[\tfrac{27}{2}\rho_{K}^{+}a_{1}^{K}
       -\tfrac{\rho_{K}^{-}}{2}(3+27\,a_{2}^{K})\Bigr]C_{1}^{1/2}(2x-1)
\nonumber\\
&\quad+\bigl(30\,\eta_{3K}+15\,\rho_{K}^{+}a_{2}^{K}
   -3\,\rho_{K}^{-}a_{1}^{K}\bigr)C_{2}^{1/2}(2x-1)
+\bigl(10\,\eta_{3K}\omega_{3K}-\tfrac{9}{2}\rho_{K}^{-}a_{2}^{K}\bigr)
   C_{3}^{1/2}(2x-1)
\nonumber\\
&\quad-3\,\eta_{3K}\omega_{3K}\,C_{4}^{1/2}(2x-1)
+\tfrac{3}{2}(\rho_{K}^{+}+\rho_{K}^{-})(1-3a_{1}^{K}+6a_{2}^{K})\ln x
\nonumber\\
&\quad+\tfrac{3}{2}(\rho_{K}^{+}-\rho_{K}^{-})(1+3a_{1}^{K}+6a_{2}^{K})\ln\xbar,
\\[4pt]
\label{eq:phisigKaon}
\phi_{K}^{\sigma}(x,\mu)
&=6x\xbar\Bigl\{
1+\tfrac{3}{2}\rho_{K}^{+}+15\,\rho_{K}^{+}a_{2}^{K}
   -\tfrac{15}{2}\rho_{K}^{-}a_{1}^{K}
\nonumber\\
&\hspace{1.2cm}+\bigl[3\,\rho_{K}^{+}a_{1}^{K}
   -\tfrac{15}{2}\rho_{K}^{-}a_{2}^{K}\bigr]C_{1}^{3/2}(2x-1)
+\bigl[5\,\eta_{3K}-\tfrac{1}{2}\eta_{3K}\omega_{3K}
   +\tfrac{3}{2}\rho_{K}^{+}a_{2}^{K}\bigr]C_{2}^{3/2}(2x-1)
\nonumber\\
&\hspace{1.2cm}+\eta_{3K}\omega_{3K}\,C_{3}^{3/2}(2x-1)\Bigr\}
+9x\xbar(\rho_{K}^{+}+\rho_{K}^{-})(1-3a_{1}^{K}+6a_{2}^{K})\ln x
\nonumber\\
&\quad+9x\xbar(\rho_{K}^{+}-\rho_{K}^{-})(1+3a_{1}^{K}+6a_{2}^{K})\ln\xbar,
\end{align}
with $\eta_{3K}=f_{3K}/(f_{K}\muK)$ and $\omega_{3K}$ defined in
analogy with the pion case.  As compared with the pion expansions
\eqref{eq:phipPion}--\eqref{eq:phisigPion}, the kaon LCDAs contain three
new structures.  The first consists of the odd-conformal-spin Gegenbauer terms
$C_{1}^{1/2,3/2}$ and $C_{3}^{1/2,3/2}$, generated by the
$G$-parity-violating combinations $\rho_{K}^{-}a_{1,2}^{K}$ and
$\omega_{3K}$ entering with $a_{1}^{K}$.  The second consists of the end point
logarithms $\ln x$ and $\ln\xbar$, which are intrinsic to the kaon and
absent for the pion (the $G$-parity-symmetric limit
$\rho_{K}^{\pm}\to 0$ removes them).  The third consists of the explicit
quark-mass corrections proportional to $\rho_{K}^{\pm}$.  Following the
conventions of Ref.~\cite{Ball:2006wn}, the constant offset
$3\rho_{K}^{+}(1+6a_{2}^{K}-9\rho_{K}^{-}a_{1}^{K})$ in
$\phi_{K}^{p}$ combines with the $\ln x,\ln\xbar$ terms so as to
preserve the canonical normalization
$\int_{0}^{1}\!\mathrm{d}x\,\phi_{K}^{p}(x,\mu)=1$ to the order
considered, whereas no such cancellation operates in $\phi_{K}^{\sigma}$,
whose normalization $\int_{0}^{1}\!\mathrm{d}x\,\phi_{K}^{\sigma}(x,\mu)$
is shifted away from unity by an amount of order
$\rho_{K}^{+}\sim 4\times 10^{-2}$~\cite{Ball:2006wn}.  The three-particle twist-3
distribution amplitude (DA) of
the kaon takes the standard form
$\phi^{K}_{3\sigma}(\alpha_{i},\mu)=360\,\alpha_{1}\alpha_{2}\alpha_{3}^{2}
[1+\lambda_{3K}(\alpha_{1}-\alpha_{2})+\tfrac{1}{2}\omega_{3K}(7\alpha_{3}-3)]$~\cite{Ball:2006wn};
it brings in the new $G$-parity-violating parameter $\lambda_{3K}$
(also vanishing in the chiral $SU(3)$ limit) but is power-suppressed by
$f_{3K}/f_{K}\sim 10^{-2}$ and is dropped in the present analysis.

All Gegenbauer moments and twist-3 parameters above depend on the
factorization scale.  We evolve them at leading order with the
Efremov-Radyushkin-Brodsky-Lepage and renormalization-group equations of
Refs.~\cite{Lepage:1980fj,Efremov:1979qk,Braun:2003rp,Ball:2006wn}.  For
the kaon, $SU(3)$ breaking induces the mixing of the twist-3 parameters
with the mass-twist-2 combinations given in Ref.~\cite{Ball:2006wn},
which we adopt.  The numerical values of all input parameters, and the discussion of
their uncertainties, are collected in Sec.~\ref{sec:inputs}.

\subsection{Light cone wave functions and helicity amplitudes}
\label{subsec:projector}

Fourier-transforming Eqs.~\eqref{eq:pionmatrix} to momentum space and
including the transverse-momentum dependence yields the standard
twist-3 spinor projector of the
pion~\cite{Beneke:2000wa,Wei:2002iu,Kroll:2018uvl,Wang:2019trq}
\begin{equation}
\label{eq:projector_pi}
M^{\pi}_{\alpha\beta}=
\frac{if_{\pi}}{4}
\left\{\slashed{p}\,\gamma_{5}\Psi_{\pi}^{\,\pi}
-\mupi\gamma_{5}\!
\left[\Psi_{\pi}^{p}
-i\sigma_{\mu\nu}\frac{p^{\mu}\bar p^{\nu}}{p\cdot\bar p}\frac{\Psi_{\pi}^{\sigma'}}{6}
+i\sigma_{\mu\nu}p^{\mu}\frac{\Psi_{\pi}^{\sigma}}{6}
\frac{\partial}{\partial k_{\perp\nu}}\right]\right\}_{\!\alpha\beta},
\end{equation}
where $\bar p^{\mu}=(p^{0},-\mathbf{p})$ is the conjugate lightlike
vector,
$\Psi_{\pi}^{\sigma'}\equiv\partial\Psi_{\pi}^{\sigma}/\partial x$, and
$\Psi_{\pi}^{i}(x,\kperp)$ are the (momentum-space) light cone wave
functions whose collinear parts coincide with $\phi_{\pi}^{i}(x)$
defined above ($i\in\{\pi,p,\sigma\}$).  We assume that the three wave
functions factorize into a longitudinal LCDA times a common
transverse-momentum profile
\begin{equation}
\label{eq:WFfactorized}
\Psi_{\pi}^{i}(x,\kperp)=\phi_{\pi}^{i}(x)\,\Sigma_{\pi}(x,\kperp),
\qquad i\in\{\pi,p,\sigma\},
\end{equation}
where the transverse profile is taken in the
Brodsky-Huang-Lepage (BHL)
form~\cite{Brodsky:1981jv,Lepage:1982gd,Jakob:1993iw,Bolz:1996wh,Wang:2019trq}
\begin{equation}
\label{eq:Gaussiankt}
\Sigma_{\pi}(x,\kperp)=\frac{16\pi^{2}\beta_{\pi}^{2}}{x\xbar}
\exp\!\left[-\frac{\beta_{\pi}^{2}\kperp^{2}}{x\xbar}\right].
\end{equation}
We stress that the BHL ansatz is not a fully factorized Gaussian
in $\kperp$ and $x$, because its harmonic-oscillator origin couples
the transverse and longitudinal degrees of
freedom through the $x\xbar$ factor in the exponent, so that the width
of the $\kperp$ distribution shrinks toward the end points
$x\to 0,1$.  This nonfactorized $x$--$\kperp$ correlation is essential
for the correct soft end point behavior and distinguishes the BHL form
from a naive $\phi(x)\exp(-\kperp^{2}/2\langle\kperp^{2}\rangle)$
ansatz.  The oscillator parameter $\beta_{\pi}$ is fixed by the
constraint on the intrinsic average transverse momentum
$\langle\kperp^{2}\rangle_{\pi}$,
\begin{equation}
\label{eq:kperpaverage}
\langle\kperp^{2}\rangle_{\pi}
=\frac{1}{2\beta_{\pi}^{2}}
\frac{\int\!\mathrm{d}x\,|\phi_{\pi}(x)|^{2}}
     {\int\!\mathrm{d}x\,|\phi_{\pi}(x)|^{2}/[x\xbar]}.
\end{equation}
In the $k_{T}$-factorization theorem the convolution is carried out in
the conjugate impact-parameter ($\mathbf{b}$) space, where
$\mathbf{b}$ is the transverse separation between the quark and the
antiquark inside the meson (conjugate to $\kperp$), and it sets the
factorization scale $\mu_{F}=1/b$ at which the soft and hard physics
are separated.  Fourier-transforming the BHL profile
\eqref{eq:Gaussiankt} to $\mathbf{b}$-space yields the clean
exponential damping factor (the hat denotes $\mathbf{b}$-space
quantities throughout)
\begin{equation}
\label{eq:bspaceSigma}
\widehat{\Sigma}_{\pi}(x,\mathbf{b})
=\int\!\frac{\mathrm{d}^{2}\kperp}{(2\pi)^{2}}
\Sigma_{\pi}(x,\kperp)e^{-i\kperp\cdot\mathbf{b}}
=4\pi\exp\!\left[-\frac{x\xbar\,\mathbf{b}^{2}}{4\beta_{\pi}^{2}}\right],
\end{equation}
which together with the Sudakov factor specified below
ensures the perturbative convergence of the convolution in the soft
end point region.  The corresponding $\mathbf{b}$-space wave functions
are $\widehat{\Psi}_{\pi}^{i}(x,\mathbf{b})=\phi_{\pi}^{i}(x)\,
\widehat{\Sigma}_{\pi}(x,\mathbf{b})$.  The kaon wave functions
$\Psi_{K}^{i}(x,\kperp)$ take the same form with
$\beta_{\pi}\to\beta_{K}$ and $\phi_{\pi}^{i}\to\phi_{K}^{i}$.

With these wave functions in hand, the $k_{T}$-factorization helicity
amplitudes for $\gam\gam\to\pizpiz$
follow~\cite{Botts:1989kf,Li:1992nu,Li:1998is,Kurimoto:2001zj,Nagashima:2002ia}
as a convolution of the two soft wave functions and the hard kernels in
the joint $(x,y;\mathbf{b}_{1},\mathbf{b}_{2})$ space
\begin{equation}
\label{eq:HelicityAmp2b}
\begin{aligned}
\mathcal{M}_{\lambda_{1}\lambda_{2}}(\gam\gam\!\to\!\pizpiz)
=&\int\!\mathrm{d}x\,\mathrm{d}y\!\int\!
   \frac{\mathrm{d}^{2}\mathbf{b}_{1}}{4\pi}
   \frac{\mathrm{d}^{2}\mathbf{b}_{2}}{4\pi}
   \!\!\!\sum_{i,j\in\{\pi,p,\sigma,\sigma'\}}\!\!\!
   \widehat{\Psi}_{\pi}^{j}(y,\mathbf{b}_{2},\mu_{F})\,
   \widehat{T}^{\lambda_{1}\lambda_{2}}_{ij}
       (x,y,Q,\theta^{*},\mathbf{b}_{1},\mathbf{b}_{2},\mu_{R})\\
&\hspace{2.8cm}\times
   \widehat{\Psi}_{\pi}^{i}(x,\mathbf{b}_{1},\mu_{F})\,
   S_{t}(x)\,S_{t}(y)\,
   \exp[-S(x,y,\mathbf{b}_{1},\mathbf{b}_{2},Q,\mu_{F},\mu_{R})],
\end{aligned}
\end{equation}
with $\widehat{\Psi}_{\pi}^{\sigma'}\equiv\partial\widehat{\Psi}_{\pi}^{\sigma}/\partial x$.
The sum over the two flavor components of the
$\pi^{0}=(u\bar u-d\bar d)/\sqrt{2}$ state, together with the photon
electric charges, is contained in the overall prefactor of the hard
kernel $\widehat{T}^{\lambda_{1}\lambda_{2}}_{ij}$ (see
Sec.~\ref{subsec:hardkernels}), so that $\mathcal{M}$ already
represents the full $\pizpiz$ amplitude.

Before discussing the resummation factors that appear in
Eq.~\eqref{eq:HelicityAmp2b}, it is useful to specify the two physical
scales that enter the convolution.  The first is the factorization
scale $\mu_{F}=1/b$, naturally identified with the inverse transverse
separation of the quark-antiquark pair, and the second is the
renormalization scale $\mu_{R}=t_{n}$, taken as the largest
characteristic invariant of the internal gluon exchange in the $n$th
diagrammatic group (defined in Sec.~\ref{subsec:hardkernels} below).
This choice of $\mu_{R}$ minimizes the higher-order QCD corrections in
the standard manner~\cite{Li:1992nu,Kroll:2018uvl}.  The strong coupling
is evaluated at one loop, matched across the charm and bottom thresholds
to the world average $\alpha_{s}(M_{Z})=0.1180$~\cite{ParticleDataGroup:2024cfk},
which corresponds to $\Lambda_{\mathrm{QCD}}^{(N_{f}=4)}\simeq
0.12~\mathrm{GeV}$.  The same scale enters the Sudakov resummation, so that
the hard coupling and the Sudakov factor are mutually consistent.  These
two scales control, respectively, the Sudakov and the threshold
resummation factors of Eq.~\eqref{eq:HelicityAmp2b}, which we now
specify in turn.

We begin with the Sudakov factor.  The exponential $\exp[-S]$ resums
to all orders the double logarithms $\alpha_{s}\ln^{2}(xQb)$ generated by
the overlap of soft and collinear gluon emission off the massless quark
lines, together with the single logarithm associated with the
renormalization-group evolution between the scales $1/b$ and $\mu_{R}$.
In the next-to-leading-logarithm
approximation~\cite{Li:1992nu,Coriano:1998ge}
\begin{equation}
\label{eq:SudakovS}
S(x,y,Q,b,t)=s(x,b,Q)+s(\xbar,b,Q)+s(y,b,Q)+s(\ybar,b,Q)
-\frac{2}{\beta_{0}}\ln\!\frac{\ln(t/\LamQCD)}{\ln[1/(b\LamQCD)]},
\end{equation}
with the universal function $s(\xi,b,Q)$ originally derived by Botts
and Sterman~\cite{Botts:1989kf} and subsequently refined to
next-to-leading-logarithmic accuracy~\cite{Li:1992nu,Jakob:1993iw}, its
explicit form being given in those references and in our earlier
work~\cite{Wang:2019trq}.  The factor $\exp[-S]$
falls off rapidly at large $b$ and effectively suppresses the
nonperturbative regime $b\gtrsim 1/\LamQCD$, providing a dynamically
generated infrared cutoff that obviates ad hoc prescriptions of the
type used in collinear analyses~\cite{Jakob:1993iw,Hsieh:2004ee}.

The Sudakov damping alone is not, however, sufficient at twist-3
accuracy.  Because the asymptotic pseudoscalar DA
$\phi_{p}^{\mathrm{as}}=1$ does not vanish at the partonic end points,
the two-parton twist-3 contributions generate additional logarithmic
end point enhancements $\alpha_{s}\ln^{2}x$.
These are resummed into a jet function $S_{t}(x)$ via the threshold
resummation~\cite{Li:1998is,Kurimoto:2001zj,Li:2001ay}
\begin{equation}
\label{eq:St}
S_{t}(x,Q)=\frac{2^{1+2c}\Gamma(\tfrac{3}{2}+c)}{\sqrt{\pi}\,\Gamma(1+c)}
\bigl[x\xbar\bigr]^{c},
\qquad
c(Q)=\min\bigl\{\,0.04Q^{2}-0.51Q+1.87,\,1\bigr\},
\end{equation}
where the parabolic parametrization of the parameter $c$ is taken
from the Li-Mishima fit of the pion transition form
factor~\cite{Li:2009pr}.  Because $S_{t}(x)\to 0$ as $x\to 0,1$, the
remaining soft-end point contributions are damped sufficiently to keep
the perturbative calculation self-consistent down to
$Q\sim 2~\mathrm{GeV}$.

With the Sudakov and the threshold resummation factors specified, we
can now reduce the two-impact-parameter convolution
\eqref{eq:HelicityAmp2b} to a single $b$-integration.  Starting from
the Fourier representation of the hard kernel in $\mathbf{b}$-space
\begin{equation}
\label{eq:Tb}
\widehat{T}^{\lambda_{1}\lambda_{2}}_{ij}
   (x,y,Q,\theta^{*},\mathbf{b}_{1},\mathbf{b}_{2},\mu_{R})
=\!\int\!\frac{\mathrm{d}^{2}\kperpi{1}}{(2\pi)^{2}}
   \frac{\mathrm{d}^{2}\kperpi{2}}{(2\pi)^{2}}
   T^{\lambda_{1}\lambda_{2}}_{ij}
       (x,y,Q,\theta^{*},\kperpi{1},\kperpi{2},\mu_{R})\,
   e^{-i\kperpi{1}\cdot\mathbf{b}_{1}-i\kperpi{2}\cdot\mathbf{b}_{2}}.
\end{equation}
The parity invariance of the underlying $\gam\gam\to q\bar q q\bar q$
amplitude implies
\begin{equation}
\label{eq:parity}
\widehat{T}^{++}_{ij}=\widehat{T}^{--}_{ij},
\qquad
\widehat{T}^{+-}_{ij}=\widehat{T}^{-+}_{ij},
\end{equation}
so that only two independent helicity components must be retained.
Within the standard small-$x$ kinematic
hierarchy~\cite{Li:2010nn,Hu:2012cp}
\[
Q^{2}\gg xQ^{2},\xbar Q^{2},yQ^{2},\ybar Q^{2}
   \gg\xbar yQ^{2},x\ybar Q^{2},\kperpi{1}^{2},\kperpi{2}^{2},
\]
the transverse-momentum dependence is dropped from the quark propagators,
where it is subleading, and retained only in the internal gluon
propagator, where it regulates the end point singularity.  The hard
kernel then localizes in $\mathbf{b}$-space at
$\mathbf{b}_{1}+\mathbf{b}_{2}=0$, so that
\begin{equation}
\widehat{T}^{\lambda_{1}\lambda_{2}}_{ij}
   (x,y,Q,\theta^{*},\mathbf{b}_{1},\mathbf{b}_{2},\mu_{R})\;\propto\;
\delta^{(2)}(\mathbf{b}_{1}+\mathbf{b}_{2}).
\end{equation}
After performing the trivial azimuthal angle integration the helicity
amplitude takes the compact one-$b$ form
\begin{equation}
\label{eq:HelicityAmp1b}
\begin{aligned}
\mathcal{M}_{\lambda_{1}\lambda_{2}}(\gam\gam\!\to\!\pizpiz)=&
\int_{0}^{1}\!\mathrm{d}x\!\int_{0}^{1}\!\mathrm{d}y\!
   \int_{0}^{\infty}\!\frac{b\,\mathrm{d}b}{(4\pi)^{2}}
   \!\!\sum_{i,j\in\{\pi,p,\sigma,\sigma'\}}\!\!
   \sum_{n\in\{a,b,c,d\}}\!\!
   \widehat{\Psi}_{\pi}^{j}(y,b,1/b)\,
   \TildeT^{\lambda_{1}\lambda_{2}}_{nij}(x,y,Q,\theta^{*},b,t_{n})\\
&\times\widehat{\Psi}_{\pi}^{i}(x,b,1/b)\,
   S_{t}(x)\,S_{t}(y)\,\exp[-S(x,y,Q,b,t_{n})],
\end{aligned}
\end{equation}
where $\TildeT^{\lambda_{1}\lambda_{2}}_{nij}$ are the partial hard
kernels associated with the four diagrammatic groups $n=a,b,c,d$
displayed in Fig.~\ref{fig:feyn}.  Each group corresponds to a
topologically distinct virtuality of the internal gluon line, and the
remaining diagrams of each group are generated by permutations of the
gluon and quark lines.  The upper $b$-integration limit is taken at
$b_{\max}\sim 4.5~\mathrm{GeV}^{-1}$ in the numerical analysis to
avoid the $\alpha_{s}$ Landau pole, while the Sudakov damping
\eqref{eq:SudakovS} makes the integrand fall well below this scale.

\subsection{Hard kernels for \texorpdfstring{$\gam\gam\to\pizpiz$}{gamma gamma to pi0 pi0}}
\label{subsec:hardkernels}

For each group $n\in\{a,b,c,d\}$ the partial hard kernel factorizes as
\begin{equation}
\label{eq:Thard}
\TildeT^{\lambda_{1}\lambda_{2}}_{nij}(x,y,Q,\theta^{*},b,t_{n})
\propto \kappa_{m}\,
\bigl[\mathcal{R}^{\lambda_{1}\lambda_{2}}_{n,ij}(x,y,\theta^{*})\bigr]\,
\mathrm{F}(s_{n},b)\,\mathrm{B}_{n,ij}(pb),
\end{equation}
where $\mathcal{R}^{\lambda_{1}\lambda_{2}}_{n,ij}$ is a rational
function of the partonic fractions and the angular variables $s,c$
alone (coming from the Dirac trace), and the prefactors $\kappa_{1,2}$
collect the electromagnetic and strong couplings together with the
charge-flavor structure of the $\pi^{0}$.  The latter follows from
$\pi^{0}=(u\bar u-d\bar d)/\sqrt{2}$, since the two photons couple to a
common quark line of flavor $q$ and the coherent sum over the $u$ and
$d$ components---weighted by the squared flavor-projection
$(1/\sqrt{2})^{2}$ and the quark charges---yields the charge-flavor
factor $(e_{u}^{2}+e_{d}^{2})/2$ that is common to all four
diagrammatic groups.  Hence the electric charge $e_{q}$ is already
absorbed into the prefactors, which read
\begin{equation}
\label{eq:kappas}
\kappa_{1}=\frac{e_{u}^{2}+e_{d}^{2}}{2}\,
   \pi^{2}\alpha_{\mathrm{em}}\,\alpha_{s}(t_{n})\,f_{\pi}^{2},
\qquad
\kappa_{2}=\frac{\mupi^{2}}{\omega^{2}}\,\kappa_{1},
\end{equation}
with $e_{u}=+\tfrac{2}{3}$ and $e_{d}=-\tfrac{1}{3}$.  The factor
$\kappa_{2}/\kappa_{1}=\mupi^{2}/\omega^{2}=4\mupi^{2}/Q^{2}$ is the
chiral-enhancement parameter discussed in Sec.~\ref{sec:intro}.  In the
Belle window $Q\sim 2$--$4~\mathrm{GeV}$ this ratio is of order
$1.8$--$7$, so the twist-3 corrections are not parametrically
suppressed.  In particular, the chiral masses $\mupi$ and $\muK$ are of
order the hard scale in this window, which is the physical reason why
the two-parton twist-3 contributions can become comparable to or even
exceed the leading-twist ones.  The allowed range of $\mu_{M}$ is
therefore directly tied to the importance of the twist-3 sector and is
discussed in Sec.~\ref{sec:inputs}.
The four group-specific invariants of the internal gluon are
\begin{equation}
\label{eq:sn}
\begin{aligned}
s_{a}&=4\xbar y\,\omega^{2},
&\qquad s_{b}&=4x\ybar\,\omega^{2},\\
s_{c}&=4(x-c^{2})(\ybar-s^{2})\,\omega^{2},
&\qquad s_{d}&=4(\xbar-c^{2})(y-s^{2})\,\omega^{2},
\end{aligned}
\end{equation}
and the corresponding renormalization scales
\begin{equation}
\label{eq:tn}
\begin{aligned}
t_{a}&=\max\!\bigl\{2\omega\sqrt{\xbar y},\,1/b\bigr\},
&\qquad t_{b}&=\max\!\bigl\{2\omega\sqrt{x\ybar},\,1/b\bigr\},\\
t_{c}&=\max\!\bigl\{2\omega\sqrt{c^{2}\xbar\ybar+s^{2}xy},\,1/b\bigr\},
&\qquad t_{d}&=\max\!\bigl\{2\omega\sqrt{s^{2}\xbar\ybar+c^{2}xy},\,1/b\bigr\}.
\end{aligned}
\end{equation}
Note that $s_{a},s_{b}$ are sign-definite (always positive in the
physical region), whereas $s_{c},s_{d}$ change sign as a function of
$x,y,\theta^{*}$, signalling the timelike/spacelike character of the
exchanged gluon in the cross-quark diagrams.  The Fourier-transformed
propagator function reads accordingly
\begin{equation}
\label{eq:Fsnb}
\mathrm{F}(s_{n},b)=
\begin{cases}
\;2\pi\,\Hone(\sqrt{s_{n}}\,b),       & s_{n}>0,\\[3pt]
-4i\,\Kzero(\sqrt{-s_{n}}\,b),        & s_{n}<0,
\end{cases}
\end{equation}
with $\Hone$ the Hankel function of the first kind and $\Kzero$ the
modified Bessel function of the second kind.  The Bessel factor
$\mathrm{B}_{n,ij}(pb)$, with $p=Qsc$, is $\Jzero(pb)$, $\Jone(pb)$ or
$\Jtwo(pb)$ depending on the helicity and spin structure of the
particular diagram (see the explicit expressions below).  Its physical
origin is the Fourier transform of the additional photon-side
transverse momentum carried by the cross-quark exchange diagrams of
groups $c$ and $d$---it is therefore intrinsically a
transverse-momentum effect that is absent in the collinear LT
calculation.

We now list the explicit kernels for the independent groups $a$ and
$c$.  Those of groups $b$ and $d$ follow from the discrete symmetries
given below.  The twist-2 hard kernels, built from the
$\phi^{\,\pi}\otimes\phi^{\,\pi}$ convolution with the
chiral-enhancement-free prefactor $\kappa_{1}$,
read~\cite{Wang:2015mod,Wang:2019trq}
\begin{align}
\label{eq:THpipi}
\TildeT^{++}_{a\pi\pi}&=
\frac{4}{9}\kappa_{1}\,
\frac{xy+\xbar\ybar}{s^{2}c^{2}\,x\ybar}\,\mathrm{F}(s_{a},b),
\\[2pt]
\TildeT^{+-}_{a\pi\pi}&=
\frac{4}{9}\kappa_{1}
\left[4+\frac{s^{2}(\xbar y+x\ybar)}{c^{2}x\ybar}
       +\frac{c^{2}(\xbar y+x\ybar)}{s^{2}x\ybar}\right]\mathrm{F}(s_{a},b),
\nonumber\\[2pt]
\TildeT^{++}_{c\pi\pi}&=
\frac{4}{9}\kappa_{1}
\left[\frac{xy+\xbar\ybar}{s^{2}xy}+\frac{xy+\xbar\ybar}{c^{2}\xbar\ybar}\right]
\Jzero(pb)\,\mathrm{F}(s_{c},b),
\nonumber\\[2pt]
\TildeT^{+-}_{c\pi\pi}&=
-\frac{4}{9}\kappa_{1}
\left[4-\frac{s^{2}(\xbar y+x\ybar)}{c^{2}\xbar\ybar}
       -\frac{c^{2}(\xbar y+x\ybar)}{s^{2}xy}\right]\Jzero(pb)\,\mathrm{F}(s_{c},b).
\nonumber
\end{align}

The twist-3 hard kernels are most naturally classified by the pair of
LCDA components $(i,j)$ with $i,j\in\{p,\sigma,\sigma'\}$ that they
multiply.  At twist-3 accuracy and in the asymptotic conformal
expansion there are nine independent kernels for the helicity-conserving
$(++)$ amplitude---$(p,p)$, $(p,\sigma')$, $(\sigma',p)$,
$(p,\sigma)$, $(\sigma,p)$, $(\sigma',\sigma')$, $(\sigma',\sigma)$,
$(\sigma,\sigma')$---plus an additional
$(\sigma,\sigma)$ structure that contributes only to the
helicity-flipping $(+-)$ amplitude (it carries
$\Jtwo(pb)$, which vanishes by parity in the $(++)$ projection).
The $(p,p)$ and $(p,\sigma')$ kernels of group $a$ read
\begin{align}
\label{eq:THpp}
\TildeT^{++}_{app}&=
-\frac{2}{9}\kappa_{2}
\left[2-\frac{1-\xbar y}{s^{2}x\ybar}-\frac{1-\xbar y}{c^{2}x\ybar}\right]\mathrm{F}(s_{a},b),
\\[2pt]
\TildeT^{+-}_{app}&=
\frac{2}{9}\kappa_{2}
\left[\frac{1}{s^{2}c^{2}}-\frac{2(1-x\ybar)}{\xbar y}\right]\mathrm{F}(s_{a},b),
\nonumber\\[2pt]
\TildeT^{++}_{ap\sigma'}&=
\frac{1}{27}\kappa_{2}\!
\left[\frac{1}{s^{2}c^{2}}-\frac{2}{x}\right]\mathrm{F}(s_{a},b),
\nonumber\\[2pt]
\TildeT^{+-}_{ap\sigma'}&=
\frac{1}{27}\kappa_{2}\!
\left[\frac{\xbar^{2}}{s^{2}c^{2}x\xbar}-\frac{2}{x\xbar}\right]\mathrm{F}(s_{a},b).
\nonumber
\end{align}
The $(\sigma',\sigma')$ kernels of group $a$ are
\begin{align}
\label{eq:THpp2}
\TildeT^{++}_{a\sigma'\sigma'}&=
-\frac{1}{162}\kappa_{2}\!
\left[\frac{s^{2}(1-\xbar y)}{c^{2}x\ybar}+\frac{c^{2}(1-\xbar y)}{s^{2}x\ybar}\right]
\mathrm{F}(s_{a},b),
\\[2pt]
\TildeT^{+-}_{a\sigma'\sigma'}&=
-\frac{1}{162}\kappa_{2}\!
\left[\frac{s^{2}(1+\xbar y)}{c^{2}x\ybar}+\frac{c^{2}(1+\xbar y)}{s^{2}x\ybar}\right]
\mathrm{F}(s_{a},b).
\nonumber
\end{align}
The corresponding $(p,p)$ and $(p,\sigma')$ kernels of group $c$ read
\begin{align}
\label{eq:THccrosspart}
\TildeT^{++}_{cpp}&=
-\frac{2}{9}\kappa_{2}\!
\left[2+\frac{1-\xbar\ybar}{s^{2}xy}+\frac{1-xy}{c^{2}\xbar\ybar}\right]
\Jzero(pb)\,\mathrm{F}(s_{c},b),
\\[2pt]
\TildeT^{+-}_{cpp}&=
-\frac{2}{9}\kappa_{2}\!
\left[\frac{1}{s^{2}c^{2}}-2\right]\Jzero(pb)\,\mathrm{F}(s_{c},b),
\nonumber\\[2pt]
\TildeT^{++}_{cp\sigma'}&=
\frac{1}{27}\kappa_{2}\!
\left[\frac{s^{2}-c^{2}}{s^{2}c^{2}}-\frac{x-\xbar}{x\xbar}\right]
\Jzero(pb)\,\mathrm{F}(s_{c},b),
\nonumber\\[2pt]
\TildeT^{+-}_{cp\sigma'}&=
\frac{1}{27}\kappa_{2}\!
\left[\frac{s^{2}x^{2}-c^{2}\xbar^{2}}{s^{2}c^{2}x\xbar}
       +\frac{x-\xbar}{x\xbar}\right]\Jzero(pb)\,\mathrm{F}(s_{c},b).
\nonumber
\end{align}
The remaining $(p,\sigma)$, $(\sigma,\sigma)$, $(\sigma,\sigma')$ and
$(\sigma',\sigma')$ kernels appear only in groups $c,d$, and carry the
Bessel factors $\Jone(pb)$ and $\Jtwo(pb)$ generated by the
$\partial/\partial\kperpi{\nu}$ derivative term of the
projector~\eqref{eq:projector_pi} acting through the gluon transverse
momentum.  The $(p,\sigma)$ and $(\sigma,\sigma)$ kernels read
\begin{align}
\label{eq:THccrosspart2}
\TildeT^{++}_{cp\sigma}&=
-\frac{1}{27}\kappa_{2}\,
\frac{\omega b\,(s^{2}x+c^{2}\xbar)}{sc\,x\xbar}\,\Jone(pb)\,\mathrm{F}(s_{c},b),
\\[2pt]
\TildeT^{+-}_{cp\sigma}&=
-\frac{1}{27}\kappa_{2}\,
\frac{\omega b\,(s^{2}\xbar+c^{2}x+1)}{sc\,x\xbar}\,\Jone(pb)\,\mathrm{F}(s_{c},b),
\nonumber\\[2pt]
\TildeT^{+-}_{c\sigma\sigma}&=
-\frac{1}{81}\kappa_{2}\!
\left[\frac{\omega^{2}b^{2}}{xy}+\frac{\omega^{2}b^{2}}{\xbar\ybar}\right]
\Jtwo(pb)\,\mathrm{F}(s_{c},b),
\nonumber
\end{align}
while the $(\sigma,\sigma')$ and $(\sigma',\sigma')$ kernels are
\begin{align}
\label{eq:THccrosspart3}
\TildeT^{++}_{c\sigma\sigma'}&=
\frac{1}{162}\kappa_{2}\!
\left[\frac{\omega b\,s}{c\,\ybar}-\frac{\omega b\,c}{s\,y}\right]
\Jone(pb)\,\mathrm{F}(s_{c},b),
\\[2pt]
\TildeT^{+-}_{c\sigma\sigma'}&=
\frac{1}{162}\kappa_{2}\!
\left[\frac{\omega b\,s(1+x)}{c\,\xbar\ybar}
       -\frac{\omega b\,c(1+\xbar)}{s\,xy}\right]\Jone(pb)\,\mathrm{F}(s_{c},b),
\nonumber\\[2pt]
\TildeT^{++}_{c\sigma'\sigma'}&=
-\frac{1}{162}\kappa_{2}\!
\left[\frac{s^{2}(1-xy)}{c^{2}\xbar\ybar}+\frac{c^{2}(1-\xbar\ybar)}{s^{2}xy}\right]
\Jzero(pb)\,\mathrm{F}(s_{c},b),
\nonumber\\[2pt]
\TildeT^{+-}_{c\sigma'\sigma'}&=
-\frac{1}{162}\kappa_{2}\!
\left[\frac{s^{2}(1+xy)}{c^{2}\xbar\ybar}+\frac{c^{2}(1+\xbar\ybar)}{s^{2}xy}\right]
\Jzero(pb)\,\mathrm{F}(s_{c},b).
\nonumber
\end{align}

The kernels of groups $b$ and $d$ are obtained from those of groups
$a$ and $c$, respectively, by discrete operations of the kinematic
variables that reflect the exchange symmetries of the partonic
amplitude
\begin{equation}
\label{eq:abcdrelations}
\begin{aligned}
\TildeT^{\lambda_{1}\lambda_{2}}_{bii}&=
\TildeT^{\lambda_{1}\lambda_{2}}_{aii}
   (s\leftrightarrow c,\,x\leftrightarrow y),
&
\TildeT^{\lambda_{1}\lambda_{2}}_{b\sigma'p}&=
\TildeT^{\lambda_{1}\lambda_{2}}_{ap\sigma'}
   (s\leftrightarrow c,\,x\leftrightarrow y),\\[2pt]
\TildeT^{\lambda_{1}\lambda_{2}}_{dii}&=
\TildeT^{\lambda_{1}\lambda_{2}}_{cii}
   (x\leftrightarrow\xbar,\,y\leftrightarrow\ybar),
&
\TildeT^{\lambda_{1}\lambda_{2}}_{dp\sigma}&=
\TildeT^{\lambda_{1}\lambda_{2}}_{cp\sigma}
   (x\leftrightarrow\xbar,\,y\leftrightarrow\ybar),\\[2pt]
\TildeT^{\lambda_{1}\lambda_{2}}_{dp\sigma'}&=
-\TildeT^{\lambda_{1}\lambda_{2}}_{cp\sigma'}
   (x\leftrightarrow\xbar,\,y\leftrightarrow\ybar),
&
\TildeT^{\lambda_{1}\lambda_{2}}_{d\sigma\sigma'}&=
-\TildeT^{\lambda_{1}\lambda_{2}}_{c\sigma\sigma'}
   (x\leftrightarrow\xbar,\,y\leftrightarrow\ybar).
\end{aligned}
\end{equation}
In addition, the Bose symmetry of the identical-meson final state and
the parity of the LCDA convolutions imply
\begin{equation}
\label{eq:bosesym}
\begin{aligned}
\TildeT^{\lambda_{1}\lambda_{2}}_{n\sigma p}&=
\TildeT^{\lambda_{1}\lambda_{2}}_{np\sigma}
   (s\leftrightarrow c,\,x\leftrightarrow\ybar),\\[2pt]
\TildeT^{\lambda_{1}\lambda_{2}}_{n\sigma'p}&=
-\TildeT^{\lambda_{1}\lambda_{2}}_{np\sigma'}
   (s\leftrightarrow c,\,x\leftrightarrow\ybar),\\[2pt]
\TildeT^{\lambda_{1}\lambda_{2}}_{n\sigma'\sigma}&=
-\TildeT^{\lambda_{1}\lambda_{2}}_{n\sigma\sigma'}
   (s\leftrightarrow c,\,x\leftrightarrow\ybar).
\end{aligned}
\end{equation}
Here the $a\leftrightarrow b$ relation implements the exchange of the two
final-state mesons ($x\leftrightarrow y$ and $s\leftrightarrow c$,
corresponding to $\theta^{*}\to\pi-\theta^{*}$), the $c\leftrightarrow d$
relation the quark-antiquark interchange inside each meson
($x\to\xbar$, $y\to\ybar$, with the relative signs set by the parity of
the corresponding bilinear projectors), and the Bose
symmetry~\eqref{eq:bosesym} links the cross-LCDA kernels to the direct
ones within each group.  Together these relations reduce the number of
independent kernels from the formal $40$ to $8$ for group $a$ (or $c$),
which also serves as an internal consistency check on the calculation.

The kernels exhibit several physical features.  In the chiral limit the
photon couples to a $q\bar q$ pair of opposite helicities, so the twist-2
amplitude~\eqref{eq:THpipi} favors the helicity-conserving $(++)$
configuration, whose universal denominator
$s^{2}c^{2}=\sin^{2}(\theta^{*}/2)\cos^{2}(\theta^{*}/2)$ vanishes only at
$\theta^{*}\to 0,\pi$, so that within the Belle range
$|\cos\theta^{*}|<0.6$~\cite{Uehara:2009cka,Uehara:2013mbo} the $(++)$
kernel is moderately enhanced over the $(+-)$ one.  The
chirality-flipping twist-3 projectors of Eq.~\eqref{eq:projector_pi}
populate both helicity channels at the same order in $\mu_{M}/Q$, with the
$(\sigma,\sigma)$ kernel
$\TildeT^{+-}_{c\sigma\sigma}\propto\Jtwo(pb)$ feeding the helicity-flip
amplitude alone---the quark-mass-induced chirality flip being the
mechanism by which twist-3 reaches helicity channels forbidden by massless
QCD at leading twist.  The Bessel factors $\Jone(pb),\Jtwo(pb)$ in the
$(p,\sigma)$, $(\sigma,\sigma)$ and $(\sigma,\sigma')$ kernels of groups
$c,d$ arise from the $\partial/\partial\kperp$ term of the
projector~\eqref{eq:projector_pi} acting through the photon-side
transverse momentum $pb=Qsc\,b$, and are genuine transverse-momentum
effects with no collinear analog, their end point enhancements being
cured at large $b$ by the Gaussian damping~\eqref{eq:bspaceSigma} and the
Sudakov factor.  Finally, since $s_{a},s_{b}>0$ throughout the physical
region while $s_{c},s_{d}$ change sign, groups $a,b$ involve only the
Hankel piece of Eq.~\eqref{eq:Fsnb} (timelike gluon exchange) whereas the
cross-quark groups $c,d$ carry both $\Hone$ and $\Kzero$ contributions,
reflecting the two color-flow topologies of the
$\gam\gam\to q\bar q q\bar q$ amplitude.

A noteworthy consequence is that the two-parton twist-3 contribution can
exceed the twist-2 one in the intermediate-$Q$ window.  The twist-3
kernels are folded with the asymptotic constant $\phi_{p}^{\mathrm{as}}=1$,
which does not vanish at the end points and renders the convolution
power-enhanced near $x=0,1$, in contrast to the twist-2 case folded with
$\phi_{2}\sim 6x\xbar$.  Although the twist-3 prefactor
$\kappa_{2}/\kappa_{1}=4\mupi^{2}/Q^{2}$ is formally
$1/Q^{2}$-suppressed, the chiral enhancement $\mupi\sim 2.6~\mathrm{GeV}$
together with this end point behavior makes the twist-3 contribution
exceed the twist-2 one for $Q\lesssim 3.5~\mathrm{GeV}$, as noted by
Gorsky~\cite{Gorsky:1989ev} and quantified for the charged channels in
Ref.~\cite{Wang:2019trq}.

\subsection{Extension to \texorpdfstring{$\gam\gam\to K_{S}^{0}K_{S}^{0}$}{gamma gamma to K0S K0S}}
\label{subsec:KSKS}

The short-lived neutral kaon $K_{S}^{0}$ is to an excellent
approximation [neglecting the $CP$-violating mixing parameter
$|\epsilon_{K}|\sim 2.2\times 10^{-3}$, whose effect on the cross
section is of $\mathcal{O}(10^{-6})$] the $CP$-even combination of the
strangeness eigenstates,
\begin{equation}
\label{eq:KSdef}
|K^{0}_{S}\rangle=\frac{1}{\sqrt{2}}
   \bigl(|K^{0}\rangle-|\bar K^{0}\rangle\bigr),\qquad
|K^{0}_{L}\rangle=\frac{1}{\sqrt{2}}
   \bigl(|K^{0}\rangle+|\bar K^{0}\rangle\bigr),
\end{equation}
in the phase convention
$CP|K^{0}\rangle=-|\bar K^{0}\rangle$.  Because strangeness is
conserved by the strong and electromagnetic interactions, the only
nonvanishing $\gam\gam$-induced amplitudes in the strangeness basis
are $\mathcal{M}(\gam\gam\to K^{0}\bar K^{0})$ and
$\mathcal{M}(\gam\gam\to\bar K^{0}K^{0})$, and these are related by
$C$-invariance of QCD+QED to the same Bose-symmetric
amplitude~\cite{Faiman:1990ya,Achasov:2007qr}.
Projecting on the $K_{S}^{0}K_{S}^{0}$ final state (a $C=P=+1$, $J=$even
state, as required by Yang's theorem~\cite{Yang:1950rg} for two real
photons) and using the vanishing of the strangeness-violating
amplitudes $\mathcal{M}(\gam\gam\to K^{0}K^{0})=\mathcal{M}(\gam\gam\to
\bar K^{0}\bar K^{0})=0$, one finds
\begin{equation}
\label{eq:Mksks}
\mathcal{M}_{\lambda_{1}\lambda_{2}}(\gam\gam\to K_{S}^{0}(p_{3})K_{S}^{0}(p_{4}))
=\mathcal{M}_{\lambda_{1}\lambda_{2}}(\gam\gam\to K^{0}(p_{3})\bar K^{0}(p_{4})).
\end{equation}
The differential cross section then carries an additional
$1/2!$ identical-particle phase-space factor for the two $K_{S}^{0}$
final mesons (which is absent in the $K^{0}\bar K^{0}$ case where the
two mesons are distinguishable by strangeness)
\begin{equation}
\label{eq:dsigmaKsConv}
\frac{d\sigma(\gam\gam\to\KsKs)}{d|\cos\theta^{*}|}
=\frac{1}{2}
\frac{d\sigma(\gam\gam\to\KzKbz)}{d|\cos\theta^{*}|},
\end{equation}
the overall factor $1/2$ being a pure quantum-mechanical
basis-rotation effect, independent of any dynamical
input~\cite{Uehara:2013mbo,Wang:2019trq}.

The $\KzKbz$ amplitude itself is computed from the same generic
$k_{T}$-factorization formula~\eqref{eq:HelicityAmp1b}, but with the
two outgoing mesons identified as $K^{0}=d\bar s$ and $\bar K^{0}=s\bar d$.
The Fourier-transformed light cone wave functions in the convolution
are constructed from the BBL conformal expansion of
$\phi_{K}^{i}(x,\mu)$ and the kaon-specific intrinsic
transverse-momentum scale $\beta_{K}$.  Because the LCDAs of $K^{0}$
and $\bar K^{0}$ are related by $x\leftrightarrow\xbar$, the kaon
amplitude reads
\begin{equation}
\label{eq:HelicityAmpKK}
\begin{aligned}
\mathcal{M}_{\lambda_{1}\lambda_{2}}(\gam\gam\!\to\!\KzKbz)=&
\int_{0}^{1}\!\mathrm{d}x\!\int_{0}^{1}\!\mathrm{d}y\!
   \int_{0}^{b_{\max}}\!\!\frac{b\,\mathrm{d}b}{(4\pi)^{2}}
   \!\!\sum_{i,j}\!\sum_{n=a,b,c,d}\!
   \widehat{\Psi}_{K}^{j}(y,b)\,
   \TildeT^{\lambda_{1}\lambda_{2}}_{K,nij}(x,y,Q,\theta^{*},b,t_{n})\\
&\times\widehat{\Psi}_{K}^{i}(\xbar,b)\,
   S_{t}(x)\,S_{t}(y)\,\exp[-S(x,y,Q,b,t_{n})].
\end{aligned}
\end{equation}
The hard kernels for the kaon share the rational structure and the
numerical color factors of the pion ones, and only the charge-flavor
prefactor differs.  They are therefore obtained from the pion kernels
of Sec.~\ref{subsec:hardkernels} by the simple replacement
$\kappa_{m}\to\kappa_{m}^{K}$ in every group $n=a,b,c,d$,
\begin{equation}
\label{eq:Ksubs}
\TildeT^{\lambda_{1}\lambda_{2}}_{K,nij}=
\TildeT^{\lambda_{1}\lambda_{2}}_{nij}\bigl(\kappa_{m}\to\kappa_{m}^{K}\bigr),
\qquad
\kappa_{1}^{K}=e_{s}^{2}\,\pi^{2}\alpha_{\mathrm{em}}\alpha_{s}(t_{n})f_{K}^{2},
\qquad
\kappa_{2}^{K}=\frac{\muK^{2}}{\omega^{2}}\kappa_{1}^{K},
\end{equation}
where the kaon charge-flavor factor is $e_{s}^{2}=\tfrac{1}{9}$.
Two features distinguish the kaon prefactor from the pion one.  First,
$f_{K}$ is here the physical neutral-kaon decay constant (and not
$f_{K}/\sqrt{2}$), since $K^{0}=d\bar s$ is a flavor eigenstate rather
than a flavor superposition, so that no $(1/\sqrt{2})^{2}$ projection weight
appears.  Second, because $e_{d}=e_{s}=-\tfrac{1}{3}$, the photon
couples with equal strength to the two valence flavors of the kaon, so
that all four diagrammatic groups carry the common charge factor
$e_{d}^{2}=e_{s}^{2}=e_{d}e_{s}=\tfrac{1}{9}$, in place of the pion weight
$(e_{u}^{2}+e_{d}^{2})/2$.

\section{Numerical analysis}
\label{sec:numerics}

\subsection{Cross sections and input parameters}
\label{subsec:dsigma}

Averaging over the initial photon polarizations, the differential
cross section for $\gam\gam\to\pizpiz$ reads
\begin{equation}
\label{eq:dsigmapi}
\frac{d\sigma(\gam\gam\to\pizpiz)}{d|\cos\theta^{*}|}
=\frac{1}{64\pi Q^{2}}\,\frac{1}{4}
\sum_{\lambda_{1},\lambda_{2}=\pm}
\bigl|\mathcal{M}_{\lambda_{1}\lambda_{2}}(\gam\gam\!\to\!\pizpiz)\bigr|^{2},
\end{equation}
where $\mathcal{M}_{\lambda_{1}\lambda_{2}}$ is the full amplitude of
Eq.~\eqref{eq:HelicityAmp1b}, whose charge-flavor prefactor
$(e_{u}^{2}+e_{d}^{2})/2$ already incorporates the coherent sum over the
$u\bar u$ and $d\bar d$ components of the $\pi^{0}$, while leaving the
chirally enhanced twist-3 corrections intact.  The differential cross
section for $\gam\gam\to\KsKs$, obtained by combining
Eqs.~\eqref{eq:dsigmaKsConv} and the $\KzKbz$ amplitude
\eqref{eq:HelicityAmpKK}, reads
\begin{equation}
\label{eq:dsigmaKsKs}
\frac{d\sigma(\gam\gam\to\KsKs)}{d|\cos\theta^{*}|}
=\frac{1}{64\pi Q^{2}}\,\frac{1}{4}
\sum_{\lambda_{1},\lambda_{2}=\pm}
\bigl|\mathcal{M}_{\lambda_{1}\lambda_{2}}(\gam\gam\!\to\!\KzKbz)\bigr|^{2}.
\end{equation}

Following the Belle analysis~\cite{Uehara:2009cka,Uehara:2013mbo}, and to
remain within the region where the perturbative treatment is reliable, we
compare with the partial cross section obtained by integrating the
differential cross section over the central angular range
$|\cos\theta^{*}|<0.6$,
\begin{equation}
\label{eq:sigma0}
\sigma_{0}(\gam\gam\to M^{0}\overline{M}^{0})
=\int_{|\cos\theta^{*}|<0.6}
\frac{d\sigma(\gam\gam\to M^{0}\overline{M}^{0})}{d|\cos\theta^{*}|}\,
d|\cos\theta^{*}|.
\end{equation}
This restriction keeps the minimum parton virtuality
$\min(|t|,|u|)>0.2\,Q^{2}$ safely above the soft scale
(Sec.~\ref{subsec:kinematics}), preserving the validity of the
perturbative calculation, and at the same time matches the angular range
over which the Belle Collaboration extracts the cross sections.

\label{sec:inputs}%
In our numerical evaluation the nonperturbative inputs---the meson decay
constants, the twist-2 Gegenbauer moments $\{a_{n}^{M}\}$, the two-parton
twist-3 parameters $\{\mu_{M},\eta_{3M},\omega_{3M},\dots\}$, and the
intrinsic transverse-momentum scales $\beta_{M}$---are set as follows.

The decay constants are taken at their physical
values~\cite{ParticleDataGroup:2024cfk},
$f_{\pi}=130.2(8)~\mathrm{MeV}$ and $f_{K}=155.7(3)~\mathrm{MeV}$.  The
strong coupling is evaluated with the one-loop running formula, using the
world-average $\alpha_{s}(M_{Z})=0.1180$~\cite{ParticleDataGroup:2024cfk}
as the single input and matching across the heavy-quark thresholds as
described in Sec.~\ref{subsec:projector}.  For
the twist-2 LCDAs we use, at the reference scale $\mu_{0}=1~\mathrm{GeV}$,
\begin{equation}
\label{eq:tw2inputs}
\begin{aligned}
a_{2}^{\pi}&=0.20\pm 0.07, & a_{4}^{\pi}&=0.10\pm 0.10,\\
a_{1}^{K}&=0.06\pm 0.03, & a_{2}^{K}&=0.25\pm 0.15,
\end{aligned}
\end{equation}
which represent a conservative average of the most recent
lattice-QCD~\cite{Bali:2019dqc} and light cone sum-rule
(LCSR)~\cite{Khodjamirian:2010vf,Khodjamirian:2017fxg,Cheng:2020vwr} determinations.  The
relatively large uncertainty on $a_{2}^{\pi}$ reflects the residual
tension between the lattice value ($a_{2}^{\pi}\sim 0.12$) and the
larger LCSR/dispersive value ($a_{2}^{\pi}\sim 0.27$), and we will
display both extremes as a measure of the DA-shape uncertainty.  The
nonzero $a_{1}^{K}$ encodes the $SU(3)$-flavor breaking generated by
the heavier strange quark, in the convention where the strange quark
occupies the $q_{2}$ position~\cite{Ball:2006wn}.

The two-parton twist-3 sector is governed by the chiral mass and the
three-particle parameters.  Following the BBL QCD
sum-rule analysis~\cite{Ball:1998je,Ball:2006wn} we take, at
$\mu=2~\mathrm{GeV}$,
\begin{equation}
\label{eq:tw3inputs}
\eta_{3\pi}=0.0127\pm 0.004,\qquad
\omega_{3\pi}=-1.1\pm 0.5,
\end{equation}
and, for the kaon, the corresponding $SU(3)$-broken values
\begin{equation}
\label{eq:tw3inputsK}
\eta_{3K}=0.0114\pm 0.004,\qquad
\omega_{3K}=-0.9\pm 0.5,\qquad
\lambda_{3K}=1.45\pm 0.35,
\end{equation}
where $\lambda_{3K}$ is the $G$-parity-violating three-particle
parameter generated by the strange-light quark-mass difference, which
vanishes in the pion case.  For the neutral kaon it enters with the
opposite sign, $\lambda_{3K}^{K^{0}}=-\lambda_{3K}$, relative to the
charged-kaon convention of Ref.~\cite{Ball:2006wn}.  The chiral
masses are fixed by Eq.~\eqref{eq:mupiGMOR}, together with the
precisely known meson masses $m_{\pi^{0}}=134.977~\mathrm{MeV}$ and
$m_{K^{0}}=497.611~\mathrm{MeV}$.  For the pion we input the well-known
quark-mass ratio $R\equiv m_{s}/m_{ud}$, with
$m_{ud}\equiv(m_{u}+m_{d})/2$, together with $m_{s}$ rather than
the individual $m_{u},m_{d}$, since $R$ is determined to much higher
precision than the single light-quark masses on the
lattice~\cite{Leutwyler:1996qg,FlavourLatticeAveragingGroupFLAG:2024oxs}.
In this scheme
\begin{equation}
\label{eq:mupiInput}
\mupi(\mu)=\frac{m_{\pi^{0}}^{2}\,R}{2\,m_{s}(\mu)},
\end{equation}
which, using the FLAG~2024 averages
$R=27.227\pm 0.081$ and
$m_{s}(2\,\mathrm{GeV})=93.46\pm 0.58~\mathrm{MeV}$~\cite{FlavourLatticeAveragingGroupFLAG:2024oxs},
gives $\mupi(2~\mathrm{GeV})=2.654\pm 0.018~\mathrm{GeV}$.  For the kaon
the relevant denominator is $m_{d}+m_{s}$ rather than $m_{u}+m_{d}$, so
the quark-mass ratio $R$ does not directly enter and we instead use the
FLAG~2024 quark masses $m_{d}=4.70(5)~\mathrm{MeV}$ and
$m_{s}=93.46(58)~\mathrm{MeV}$ in Eq.~\eqref{eq:mupiGMOR}, which gives
$\muK(2~\mathrm{GeV})=2.523\pm 0.015~\mathrm{GeV}$.  Collected
\begin{equation}
\label{eq:muMnum}
\mupi(2~\mathrm{GeV})=2.654\pm 0.018~\mathrm{GeV},\qquad
\muK(2~\mathrm{GeV})=2.523\pm 0.015~\mathrm{GeV}.
\end{equation}
These inputs are consistent with the well-known relations of chiral
perturbation theory (ChPT).  At leading order in the chiral expansion
the pion and the kaon share a common chiral mass,
$\mupi=\muK=-2\langle\bar qq\rangle/f_{\pi}^{2}$
[cf.~Eq.~\eqref{eq:mupiGMOR}], so the ratio $\muK/\mupi$ equals unity in
this limit.  Our values give $\muK/\mupi\simeq 0.95$, and the
quark-mass ratio $R$ quoted above lies close to its
leading-order ChPT estimate $2m_{K}^{2}/m_{\pi}^{2}-1\simeq 26.2$.
These few-percent deviations are precisely the $SU(3)$-breaking
corrections established in the ChPT analysis of the light-quark mass
ratios~\cite{Leutwyler:1996qg}, which confirms that the phenomenological
chiral masses used here are fully compatible with ChPT.

As emphasized in Sec.~\ref{subsec:hardkernels}, the size of $\mu_{M}$
directly controls the importance of the two-parton twist-3 sector
through the ratio $\kappa_{2}/\kappa_{1}=4\mu_{M}^{2}/Q^{2}$, which for
$\mu_{M}\simeq 2.5$--$2.7~\mathrm{GeV}$ is of order unity
throughout the Belle window $Q=2$--$4~\mathrm{GeV}$, so the twist-3
contributions are numerically as important as---and at the lower end
larger than---the leading-twist ones.  We note that older direct
QCD sum-rule extractions favored a somewhat broader range
$\mu_{\pi}\simeq 1.7$--$2.0~\mathrm{GeV}$ at $\mu=1~\mathrm{GeV}$
(equivalently $\sim 2.2$--$2.8~\mathrm{GeV}$ at $2~\mathrm{GeV}$ after
evolution).  We therefore treat $\mu_{M}$ also as a sensitivity
parameter when assessing the theoretical uncertainty of the twist-3
prediction.

Finally, the intrinsic transverse-momentum scales entering the BHL wave
function~\eqref{eq:Gaussiankt} are fixed, for the pion, by the
$\pi^{0}\to\gam\gam$ constraint, $\langle\kperp^{2}\rangle_{\pi}^{1/2}
=0.35~\mathrm{GeV}$~\cite{Lepage:1982gd,Brodsky:1981jv,Wang:2019trq},
and, for the kaon, by the recent intrinsic transverse-momentum analysis
of the pseudoscalar electromagnetic form
factors~\cite{Chai:2024tss,Chai:2025xuz}, which yields an appreciably
larger $\langle\kperp^{2}\rangle_{K}^{1/2}\simeq 0.55~\mathrm{GeV}$.
All inputs are evolved to the running factorization scale $\mu_{F}=1/b$
of the convolution by the leading-order equations of
Sec.~\ref{subsec:lcda}.

\subsection{Numerical results}
\label{subsec:totalsigma}

\begin{figure}[t]
\centering
\includegraphics[width=0.49\textwidth]{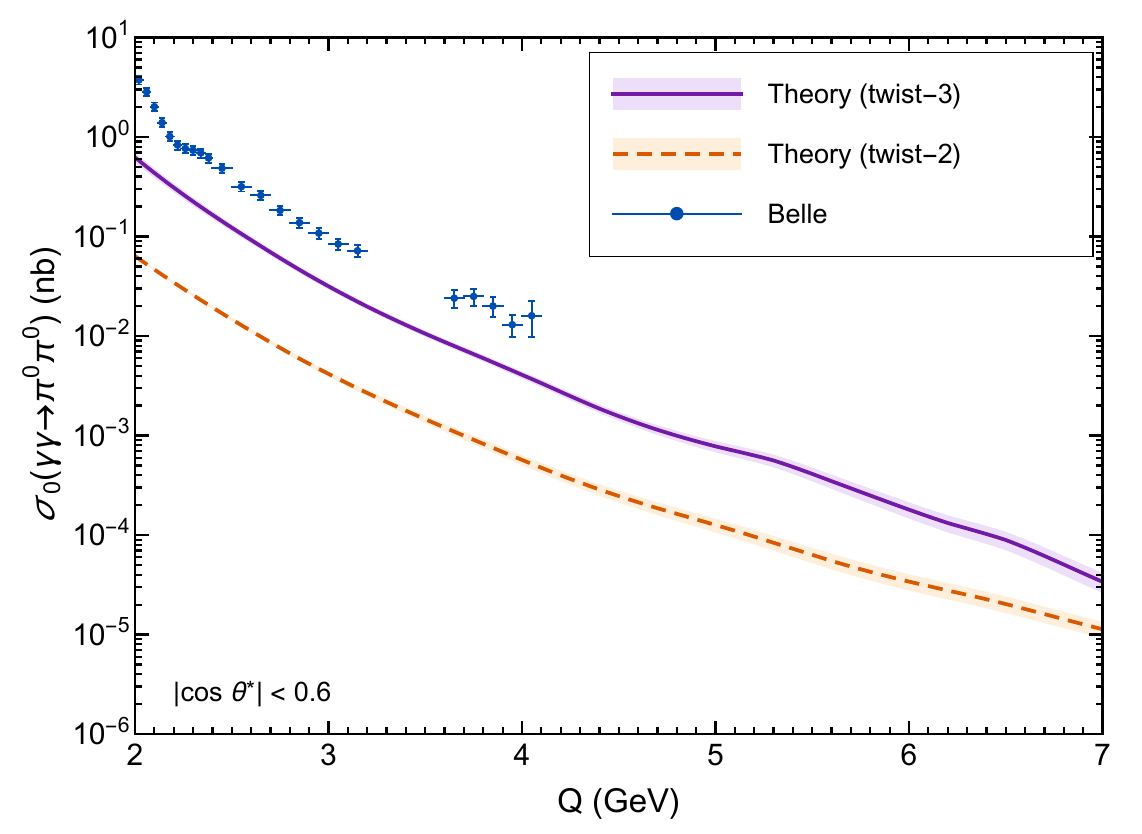}\hfill
\includegraphics[width=0.49\textwidth]{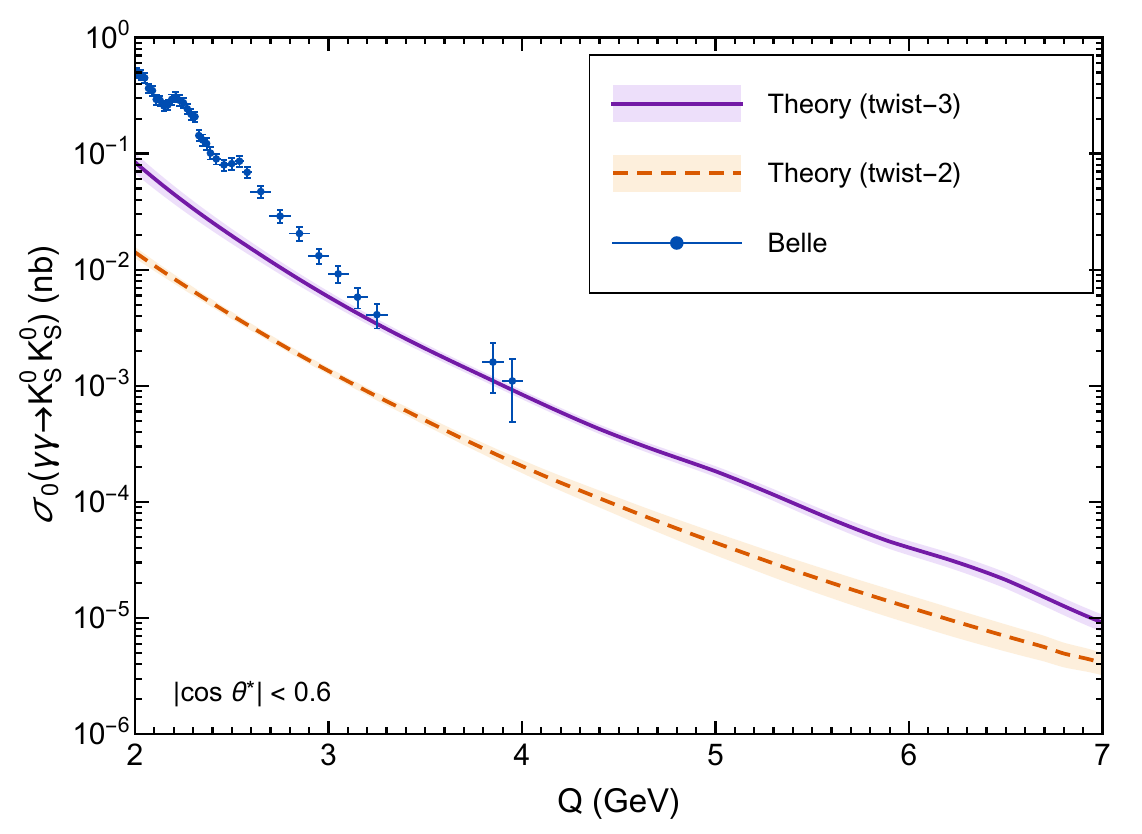}
\caption{Partial cross sections $\sigma_{0}$ ($|\cos\theta^{*}|<0.6$) for
$\gam\gam\to\pi^{0}\pi^{0}$ (left) and $\gam\gam\to K_{S}^{0}K_{S}^{0}$
(right): twist-2 (dashed) and twist-3 (solid) predictions with their
DA-input uncertainty bands, compared with the Belle
data~\cite{Uehara:2008ep,Uehara:2009cka,Chen:2006gy,Uehara:2013mbo}.}
\label{fig:sigmatot}
\end{figure}

Figure~\ref{fig:sigmatot} shows our twist-2 and twist-3 predictions for
the partial cross sections $\sigma_{0}(\gam\gam\to\pi^{0}\pi^{0})$ and
$\sigma_{0}(\gam\gam\to K_{S}^{0}K_{S}^{0})$, integrated over
$|\cos\theta^{*}|<0.6$, together with the Belle
data~\cite{Uehara:2008ep,Uehara:2009cka,Chen:2006gy,Uehara:2013mbo}.  Two
features stand out.  First, throughout the Belle window the twist-3 cross
section lies well above the twist-2 one---by up to an order of magnitude
near $Q=2~\mathrm{GeV}$---which is the direct numerical manifestation of
the chiral enhancement discussed in Secs.~\ref{sec:intro}
and~\ref{subsec:hardkernels}.  Two distinct effects combine to produce an
enhancement of this size.  The first is the chiral mass itself, through which the
twist-3 amplitude enters with the factor
$\kappa_{2}/\kappa_{1}=4\mu_{M}^{2}/Q^{2}$.  Because $\mu_{M}$ is an
order parameter of chiral symmetry breaking of order $1~\mathrm{GeV}$
rather than a current quark mass, this ratio is of order unity---about
$1.8$--$7$ across $Q=2$--$4~\mathrm{GeV}$---so the nominal $1/Q$ power
suppression of the twist-3 term is numerically defeated.  The second is
the end point behavior of the convolution, in which the asymptotic twist-3
pseudoscalar amplitude $\phi_{p}^{\mathrm{as}}=1$ does not vanish at the
partonic end points, where the hard kernels are largest.  The
twist-3 integrand therefore receives a further enhancement that the end point-suppressed
twist-2 amplitude does not.  Squaring the amplitude turns these two
$O(1)$ factors into the order-of-magnitude gap seen in
Fig.~\ref{fig:sigmatot}.  The same pattern---a twist-3 cross section
exceeding the twist-2 one by a factor ranging from several to an order of
magnitude---was found
for the charged channels $\gam\gam\to\pi^{+}\pi^{-},K^{+}K^{-}$ in the
$k_{T}$-factorization analysis of Ref.~\cite{Wang:2019trq}, so the
hierarchy is a generic feature of these reactions rather than a special
property of the neutral final states.  For the neutral channels, however,
it carries additional weight, because the leading-twist amplitude is itself
charge-suppressed (Sec.~\ref{sec:intro}).  The chirally enhanced
twist-3 and transverse-momentum contributions are therefore not merely the dominant
perturbative piece but are indispensable for any quantitative
understanding of the measured cross sections.  Their inclusion moves the
prediction much closer to the data than the leading-twist result alone.

Second, even at twist-3 the predicted cross section remains below the
measurements, by a factor of roughly three across most of the window.
This residual deficit is common to both neutral channels and, as
discussed below, is plausibly attributable to higher-order QCD
corrections.

To make the comparison quantitative we extract the effective power-law
index $n$ defined by $\sigma_{0}(Q)\sim Q^{-n}$, fitting our computed cross
sections over $Q=2$--$4~\mathrm{GeV}$ at $|\cos\theta^{*}|<0.6$.  The
results for all four channels, at twist-2 and twist-3, are collected in
Table~\ref{tab:nvalues} together with the Belle measurements.

\begin{table}[t]
\caption{Effective power-law indices $n$, defined by
$\sigma_{0}(Q)\sim Q^{-n}$ and obtained by a fit over $Q=2$--$4~\mathrm{GeV}$
at $|\cos\theta^{*}|<0.6$, for the twist-2 and twist-3 calculations,
compared with the Belle measurements.  The quoted theoretical errors are
obtained by propagating the distribution-amplitude input uncertainties
through the power-law fit.  All Belle indices refer to the same angular
range $|\cos\theta^{*}|<0.6$, and are fitted over the high-$Q$ continuum
regions of the respective experiments (excluding the charmonium window
$3.3$--$3.6~\mathrm{GeV}$).  Our index is only weakly sensitive to the fit
window---it changes by $\lesssim 0.2$ between $Q=2$--$4$ and
$Q=3$--$4~\mathrm{GeV}$---and, since the normalized angular distribution
is essentially energy independent (Fig.~\ref{fig:dsigmanorm}), also to the
angular cut, so the comparison is meaningful.}
\label{tab:nvalues}
\begin{ruledtabular}
\begin{tabular}{lccc}
Channel & $n^{\mathrm{tw2}}$ & $n^{\mathrm{tw3}}$ & $n^{\mathrm{Belle}}$ \\
\hline
$\pi^{0}\pi^{0}$   & $6.83\pm0.16$ & $7.25\pm0.14$ & $6.9\pm0.6\pm0.7$~\cite{Uehara:2009cka}\\
$K_{S}^{0}K_{S}^{0}$ & $6.14\pm0.03$ & $6.66\pm0.23$ & $10.5\pm0.6\pm0.5$~\cite{Chen:2006gy,Uehara:2013mbo}\\
$\pi^{+}\pi^{-}$   & $5.54\pm0.22$ & $7.87\pm0.07$ & $7.9\pm0.4\pm1.5$~\cite{Nakazawa:2004gu}\\
$K^{+}K^{-}$       & $4.76\pm0.48$ & $7.26\pm0.11$ & $7.3\pm0.3\pm1.5$~\cite{Nakazawa:2004gu}\\
\end{tabular}
\end{ruledtabular}
\end{table}

Three points emerge from Table~\ref{tab:nvalues}.  First, the twist-3
indices lie systematically above the twist-2 ones, the chiral $1/Q^{2}$
prefactor and the steeper end point-dominated convolution making the
twist-3 cross section fall faster with energy.  Second, the twist-3
indices are mutually similar and reproduce the measured values for three
of the four channels---the charged-channel results $n=7.87\pm0.07$ and
$n=7.26\pm0.11$ match the Belle values $7.9\pm0.4\pm1.5$ and
$7.3\pm0.3\pm1.5$, and the neutral-pion result $n=7.25\pm0.14$ agrees,
within errors, with the value $6.9\pm0.6\pm0.7$ measured at the same
angular cut $|\cos\theta^{*}|<0.6$.  All are consistent with the
$k_{T}$-factorization study of Ref.~\cite{Wang:2019trq}, the small
differences arising from our updated input parameters.  The energy
dependence of these three channels is thus set primarily by the common
hard-scattering structure.  Third, against this common pattern the neutral
kaon stands out---the measured $n_{K_{S}^{0}K_{S}^{0}}=10.5\pm0.6\pm0.5$ is
far steeper than any of the hard predictions, including our
$n^{\mathrm{tw3}}=6.66\pm0.23$.  This strong extra steepening cannot arise
from the hard mechanism alone and already points to an additional,
energy-dependent contribution---a conclusion reinforced by the ratio
analysis below.

The deficit identified above does not by itself invalidate
the angular dynamics of the hard kernels.  To test that dynamics
independently of the overall scale we consider the normalized differential
cross section $\sigma_{0}^{-1}\,d\sigma/d|\cos\theta^{*}|$, in which the
common prefactors---the chiral mass, the decay constants, and the bulk
of the DA-shape dependence---cancel between numerator and denominator.
This ratio also removes the dominant overall-normalization systematic of
the Belle measurement, so that the shape comparison is far cleaner than
the absolute one.

\begin{figure}[t]
\centering
\includegraphics[width=0.49\textwidth]{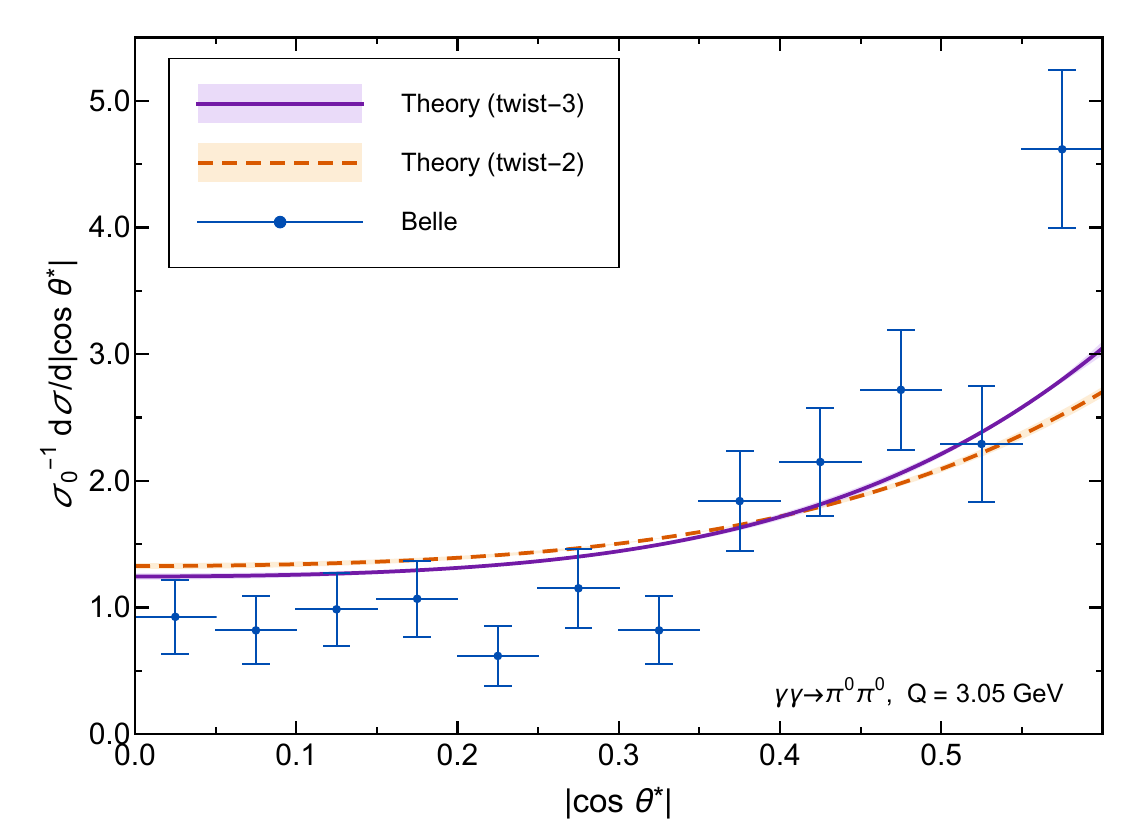}\hfill
\includegraphics[width=0.49\textwidth]{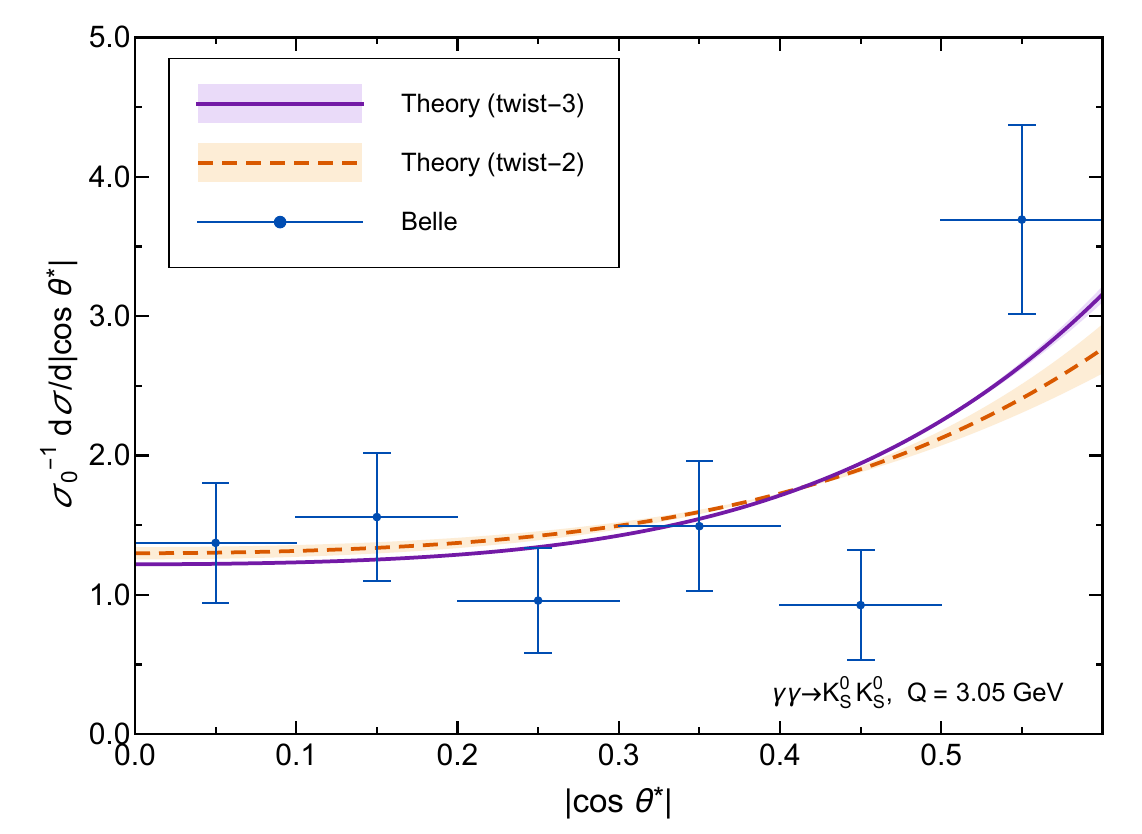}
\caption{Normalized differential cross sections
$\sigma_{0}^{-1}\,d\sigma/d|\cos\theta^{*}|$ at $Q=3.05~\mathrm{GeV}$ for
$\gam\gam\to\pi^{0}\pi^{0}$ (left) and $\gam\gam\to K_{S}^{0}K_{S}^{0}$
(right), at twist-2 and twist-3, compared with the Belle
data~\cite{Uehara:2009cka,Uehara:2013mbo}.}
\label{fig:dsigmanorm}
\end{figure}

Figure~\ref{fig:dsigmanorm} shows the normalized distributions at
$Q=3.05~\mathrm{GeV}$.  The predicted shapes agree well with the Belle
data over the full measured range $|\cos\theta^{*}|<0.6$, rising toward
larger $|\cos\theta^{*}|$ as dictated by the $s^{2}c^{2}$ denominator of
the helicity-conserving kernel modulated by the charge-suppressed angular
function of the neutral mesons.  The twist-2 and twist-3 shapes are very
similar, showing that the chiral enhancement rescales the magnitude of the
cross section without strongly distorting its angular profile.  This
agreement confirms that the helicity structure and the angular content of
the hard kernels are correctly captured, and it localizes the
theory-experiment tension in the overall normalization rather than in the
angular dynamics of the reaction.

\begin{figure}[t]
\centering
\includegraphics[width=0.49\textwidth]{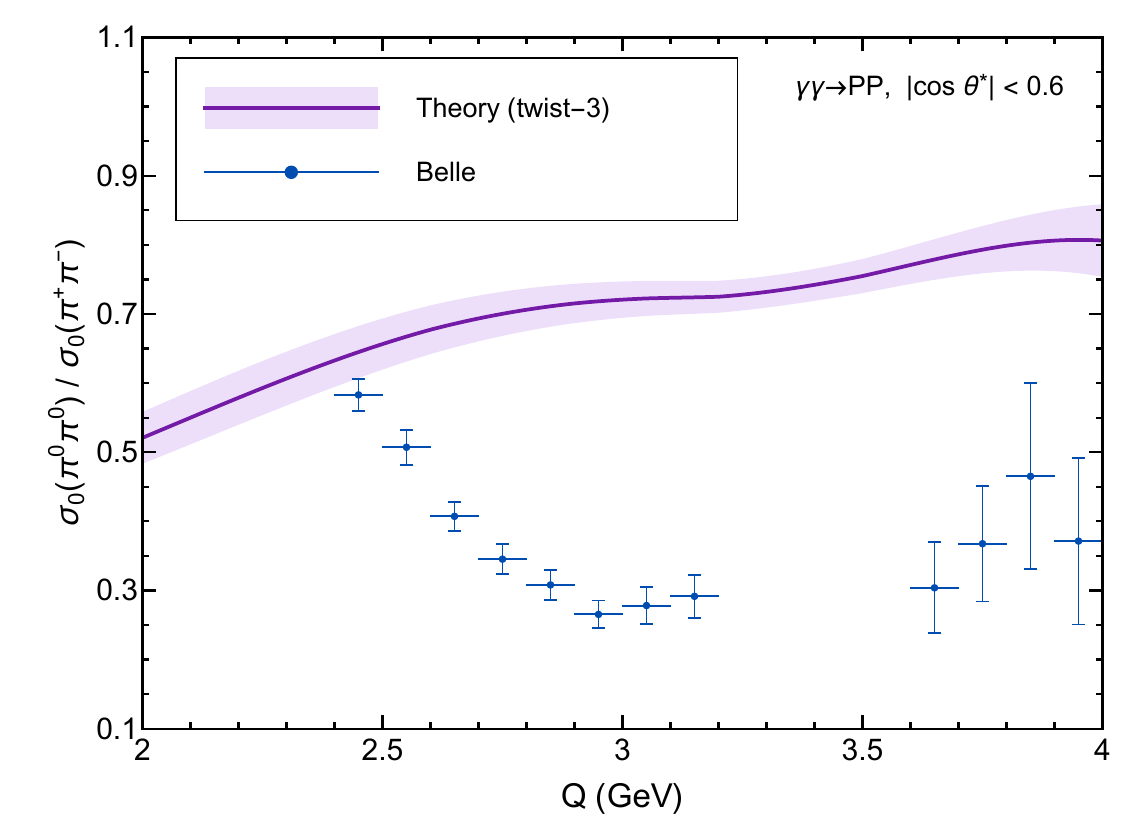}\hfill
\includegraphics[width=0.49\textwidth]{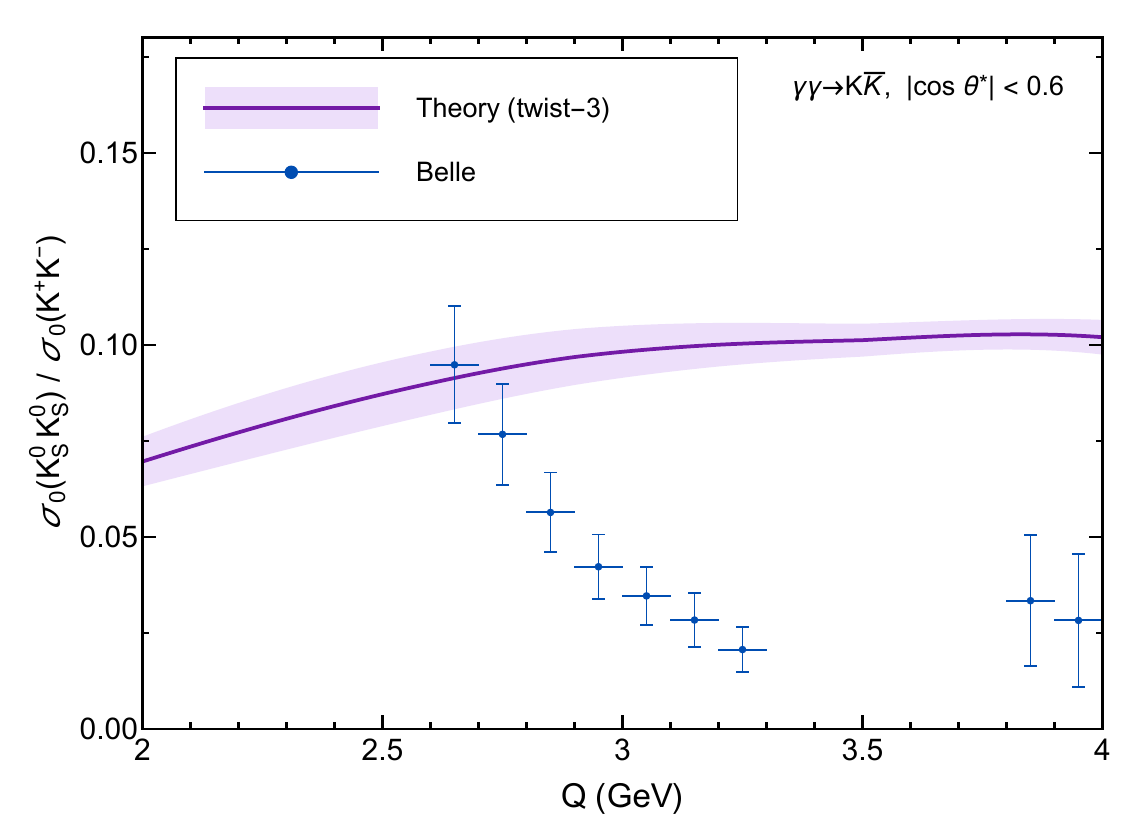}
\caption{Cross section ratios $\mathcal{R}_{\pi}$ (left) and
$\mathcal{R}_{K}$ (right) from the hard twist-3 calculation (band)
compared with the Belle
data~\cite{Uehara:2008ep,Uehara:2009cka,Chen:2006gy}.}
\label{fig:ratiohard}
\end{figure}

A still more incisive test is provided by the neutral-to-charged ratios
\begin{equation}
\label{eq:ratiodef}
\mathcal{R}_{\pi}(Q)=\frac{\sigma_{0}(\pi^{0}\pi^{0})}{\sigma_{0}(\pi^{+}\pi^{-})},
\qquad
\mathcal{R}_{K}(Q)=\frac{\sigma_{0}(K_{S}^{0}K_{S}^{0})}{\sigma_{0}(K^{+}K^{-})},
\end{equation}
in which the chiral mass, the decay constants, the higher-order $K$-factor
and most of the DA-shape uncertainty cancel to a large extent, leaving a
quantity controlled almost entirely by the charge-flavor structure and by
the relative dynamics of the neutral and charged amplitudes.
Figure~\ref{fig:ratiohard} compares the twist-3 prediction with the Belle
data.  The hard mechanism yields ratios that rise smoothly with energy---$\mathcal{R}_{\pi}$ from about $0.5$ at $Q=2~\mathrm{GeV}$ to about $0.8$
at $Q=4~\mathrm{GeV}$, and $\mathcal{R}_{K}$ from about $0.07$ to about
$0.10$---lying above the data over most of the window.  It is
instructive to set this result against the two limiting mechanisms
discussed in Sec.~\ref{sec:intro}.  Collinear leading-twist pQCD predicts
much smaller and nearly energy-independent ratios,
$\mathcal{R}_{\pi}\simeq 0.04$--$0.07$ and a comparably small
$\mathcal{R}_{K}$, whereas the soft handbag predicts
$\mathcal{R}_{\pi}\simeq 0.5$ and $\mathcal{R}_{K}\approx 2/25$, both again
almost flat in energy.  The data behave differently from all three:
$\mathcal{R}_{\pi}$ decreases from its low-$Q$ value to a minimum near
$Q\simeq 3.0~\mathrm{GeV}$ and then rises again, while $\mathcal{R}_{K}$
falls steeply by an order of magnitude across the window.  None of the
calculations---collinear leading twist, soft handbag, or the present
hard $k_{T}$-factorization result---reproduces this pattern on its own.
The monotonic rise of the hard ratios in particular is the direct
counterpart of the too-shallow power-law indices of
Table~\ref{tab:nvalues}.

That a quantity in which the leading dynamics largely cancels should vary
so strongly, and so differently in the two channels, suggests that more
than one mechanism is simultaneously important in the few-GeV region, the
observed energy dependence then arising from the interplay of these
contributions rather than from any single one.  Such a
coexistence is in fact expected on kinematic grounds, as noted in
Sec.~\ref{subsec:kinematics}, throughout the window $Q\sim 2$--$4~\mathrm{GeV}$
the minimum parton virtuality $\min(|t|,|u|)$ drops to the order of
$\LamQCD^{2}$ near the edge of the angular acceptance, so that soft
end point and handbag configurations are not parametrically separated from
the hard scattering and can feed the same amplitude coherently.  The
inability of the purely hard calculation to describe the ratios is
therefore not merely a numerical shortfall but a physical indication that
a soft contribution should be included alongside the hard one.  We pursue this
quantitatively in the following discussion.

\subsection{Discussion}
\label{subsec:ratios}

The simplest concrete realization of the soft sector identified above is
the handbag mechanism of Diehl, Kroll, and
Vogt~\cite{Diehl:2001fv,Diehl:2009yi}, in which the two photons couple to a
single quark line and the meson pair is formed through a soft annihilation
form factor.  A characteristic feature of this mechanism is that it feeds
only the opposite-helicity $(+-)$ amplitude, leaving the
helicity-conserving $(++)$ part untouched, so that it interferes
coherently with the hard $(+-)$ amplitude.  We add it at the amplitude
level
\begin{equation}
\label{eq:softhard}
\mathcal{M}^{++}=\mathcal{M}^{++}_{\rm hard},\qquad
\mathcal{M}^{+-}=\mathcal{M}^{+-}_{\rm hard}+\mathcal{A}^{+-}_{\rm soft},
\qquad
\mathcal{A}^{+-}_{\rm soft}(Q^{2},\theta^{*})=-\,\frac{16\pi\alpha}{\sin^{2}\theta^{*}}\,R_{2M}(Q^{2}),
\end{equation}
where $R_{2M}$ is the soft annihilation form factor of the relevant
channel, carrying a slowly varying modulus and an energy-dependent strong
phase.  The hard amplitude carries its own phase, generated by the
Sudakov and threshold-resummation factors and the parton-level
convolution.  We do not introduce any further relative phase between the
two amplitudes, so that the soft-to-hard phase difference is absorbed into
the phase of $R_{2M}$.  Under exact $SU(3)$ flavor symmetry the meson-pair form factors
of the different channels are related by the same charge-flavor algebra
as the hard amplitude~\cite{Diehl:2009yi}.  Writing the modulus and phase
explicitly
\begin{equation}
\label{eq:Rflavour}
\begin{aligned}
R_{\pi^{0}\pi^{0}}&=R_{\pi^{+}\pi^{-}}\equiv R_{2\pi}
   =|R_{2\pi}|\,e^{i\delta_{\pi}(Q)},\\[2pt]
R_{K^{+}K^{-}}&\equiv R_{2K}=|R_{2K}|\,e^{i\delta_{K}(Q)},\qquad
R_{K^{0}\bar K^{0}}=\tfrac{2}{5}\,|R_{2K}|\,e^{i\delta_{K}(Q)},
\end{aligned}
\end{equation}
the factor $2/5$ in the neutral-kaon form factor reflecting the smaller
charge of its valence $d$ quark relative to the $u$ quark of the charged
kaon.  Exact $SU(3)$ symmetry would in addition set
$R_{2\pi}=R_{2K}$~\cite{Diehl:2001fv}.  We retain the equality of the
moduli, taking a common value
\begin{equation}
\label{eq:Rmodulus}
Q^{2}\,|R_{2\pi}|=Q^{2}\,|R_{2K}|\simeq\mathrm{const},
\end{equation}
but, to allow for $SU(3)$-breaking effects, keep the two phases
independent, parametrizing each as
\begin{equation}
\label{eq:phaseparam}
\delta_{M}(Q)=\delta_{0}^{M}+\delta_{1}^{M}\,(Q-3~\mathrm{GeV}),
\qquad M=\pi,K.
\end{equation}

\begin{figure}[t]
\centering
\includegraphics[width=0.49\textwidth]{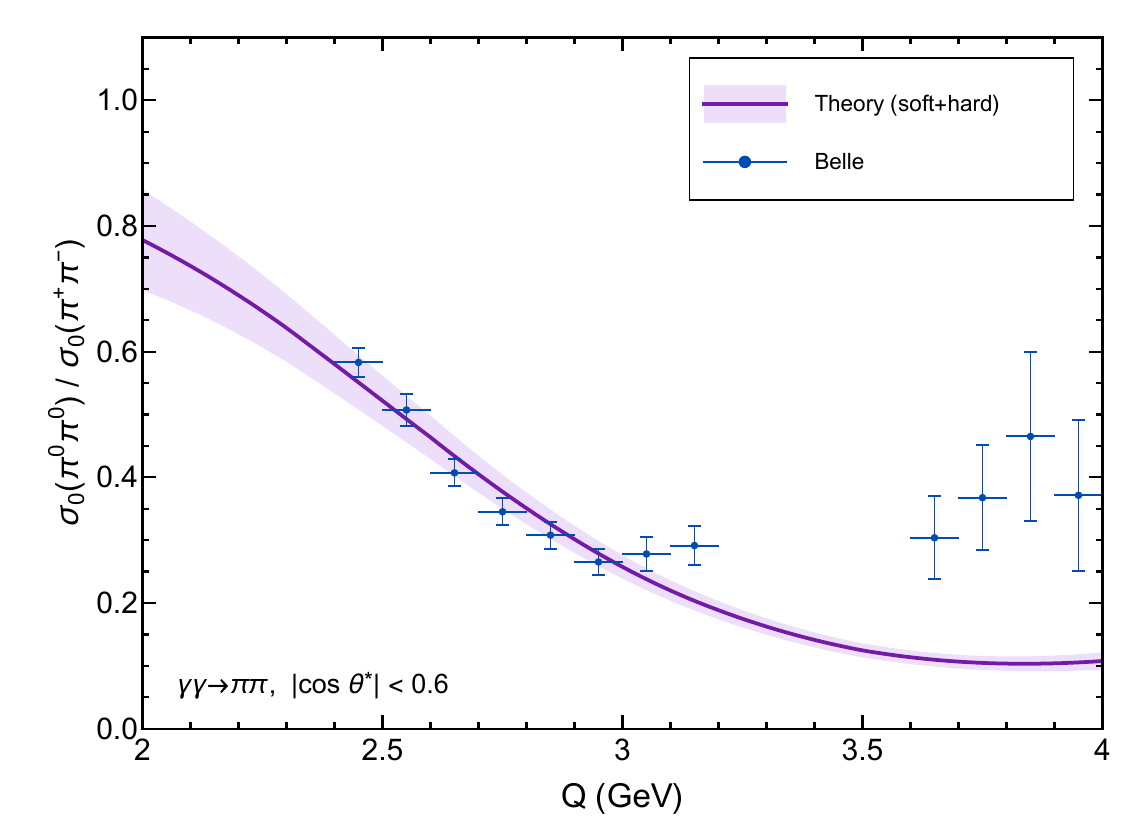}\hfill
\includegraphics[width=0.49\textwidth]{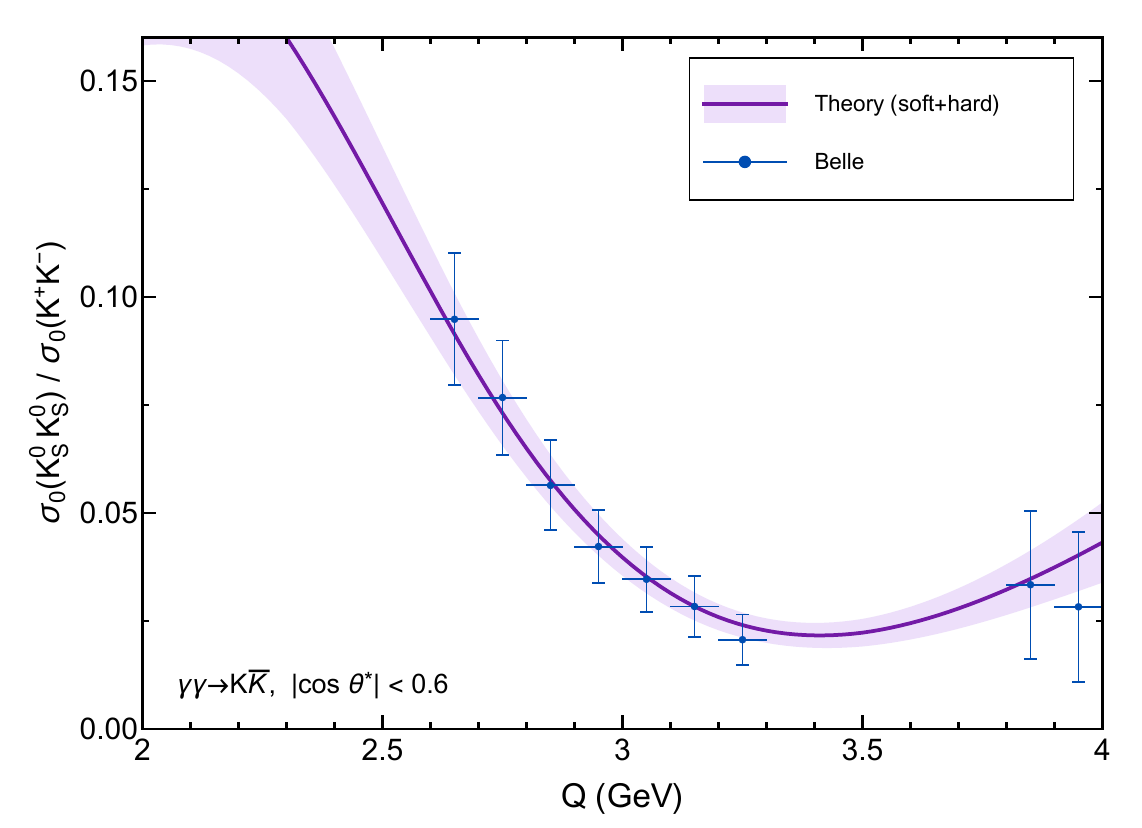}
\caption{Cross section ratios $\mathcal{R}_{\pi}$ (left) and
$\mathcal{R}_{K}$ (right) in the coherent soft-hard model (band)
compared with the Belle
data~\cite{Uehara:2008ep,Uehara:2009cka,Chen:2006gy}.}
\label{fig:ratiosofthard}
\end{figure}

Fitting the coherent soft-hard model to the two measured ratios then
yields a common modulus
\begin{equation}
\label{eq:R2Mfit}
Q^{2}\,|R_{2\pi}|=Q^{2}\,|R_{2K}|=(0.36\pm 0.11)~\mathrm{GeV}^{2},
\end{equation}
together with the phases
\begin{equation}
\label{eq:phasefit}
\begin{aligned}
\delta_{0}^{\pi}&=(246\pm 2)^{\circ}, &\quad
\delta_{1}^{\pi}&=(54\pm 6)^{\circ}/\mathrm{GeV},\\
\delta_{0}^{K}&=(247\pm 4)^{\circ}, &\quad
\delta_{1}^{K}&=(99\pm 13)^{\circ}/\mathrm{GeV}.
\end{aligned}
\end{equation}
Figure~\ref{fig:ratiosofthard} shows that with these parameters the model
reproduces the decrease of $\mathcal{R}_{\pi}$ toward its minimum and the
steep fall of $\mathcal{R}_{K}$.

Three features of this extraction carry physical information.  First, the
modulus we obtain is appreciably smaller than the value found in the
original handbag analysis:  Diehl, Kroll, and Vogt fit
$Q^{2}|R_{2\pi}|=0.75\pm 0.07~\mathrm{GeV}^{2}$ and
$Q^{2}|R_{2K}|=0.64\pm 0.04~\mathrm{GeV}^{2}$ to the two-photon
data~\cite{Diehl:2001fv}, about twice our common
$Q^{2}\,|R_{2M}|=0.36~\mathrm{GeV}^{2}$.  Their two values are close to each
other, consistent with the common modulus we adopt.  The difference in
overall size has a clear origin.  In the handbag analysis the perturbative
hard contribution is evaluated at leading twist, where it is far smaller
than the measured cross section, and is therefore neglected, the entire
cross section being assigned to the soft form factor.  Our calculation
instead retains the chirally enhanced twist-3 hard amplitude, which in the
intermediate-energy window is of comparable size to the soft one and
interferes with it coherently.  A correspondingly smaller soft form factor
then suffices to describe the data, so that the reduced $Q^{2}\,|R_{2M}|$ is a
direct consequence of the soft and hard mechanisms sharing the cross
section on an almost equal footing in this regime.

Second, the two phases have nearly identical constant parts,
$\delta_{0}^{\pi}\simeq\delta_{0}^{K}\simeq 246^{\circ}$.  Because the hard
amplitude carries a nearly $Q$-independent phase, this constant offset
fixes the overall soft-to-hard phase difference, and its near-equality for
the pion and the kaon is consistent with $SU(3)$ symmetry, under which that
phase difference is flavor independent.

Third, the energy slopes of the two phases differ markedly,
$\delta_{1}^{K}\simeq 2\,\delta_{1}^{\pi}$.  Since the moduli are taken
equal, this is where the $SU(3)$-breaking of the soft sector resides---the strange quark makes the kaon form factor phase rotate about twice as
fast with energy as the pion one.  The magnitude of these slopes is comparable
to the strong, energy-dependent variation of the soft-sector phase found
by Diehl and Kroll, whose relative form-factor phase rotates at roughly
$50$--$75^{\circ}/\mathrm{GeV}$ in the region $Q\simeq 3~\mathrm{GeV}$
where the bulk of the data lie~\cite{Diehl:2009yi}.\footnote{Diehl and
Kroll~\cite{Diehl:2009yi} parametrize the relative phase between the
valence and nonvalence soft form factors $R^{u}_{2\pi}$ and
$R^{s}_{2\pi}$ as
$\rho(s)=\pi\bigl[1+\tanh(\kappa/(s-s_{c}))\bigr]$, with
$\kappa\simeq 0.63~\mathrm{GeV}^{2}$ and $s_{c}\simeq 6~\mathrm{GeV}^{2}$,
which tends to $180^{\circ}$ at high energy.  The range quoted here is the
local slope $d\rho/dQ$ near $Q\simeq 3~\mathrm{GeV}$ ($s\simeq
9$--$10~\mathrm{GeV}^{2}$), which decreases at higher energy.  This relative
phase is not identical to our soft-to-hard phase $\delta_{M}$, but its rate
of energy variation provides the natural point of comparison.}  A
strictly common phase, $\delta_{\pi}=\delta_{K}$, cannot reproduce the two
ratios simultaneously, so the independent slopes are required by the data.

Physically, the rapid energy variation of the ratios is a coherence
effect.  As noted above, each mechanism on its own---collinear leading
twist, the soft handbag, or our hard $k_{T}$-factorization result---gives a ratio that is almost flat across $Q=2$--$4~\mathrm{GeV}$.  In this
window, however, the soft and hard $(+-)$ amplitudes are of comparable
magnitude, and their coherent sum, with a relative phase that varies with
energy, changes rapidly.  Only by retaining both contributions does the
calculation reproduce the energy dependence of the ratios observed in the
data.  This coexistence of competing soft and hard dynamics, neither of
which dominates, is what makes the few-GeV region both interesting and
difficult.

Two caveats temper this picture.  First, the soft-hard model is
introduced phenomenologically, with the modulus and phase of $R_{2M}$
fitted to the ratio data rather than computed from first principles, and
the present treatment should be regarded as a physically motivated
discussion rather than a complete dynamical account.  Second, and more
importantly, the soft contribution reshapes the ratios through the $(+-)$
channel but does not cure the absolute-normalization deficit.  The
predicted absolute cross sections remain about a factor of three below the
data even after the soft contribution is included.  This residual gap is most plausibly
attributed to the large, slowly converging higher-order QCD corrections
discussed in Sec.~\ref{sec:intro}---the NLO $K$-factor of order $1.9$
found by Duplan\v{c}i\'c and Ni\v{z}i\'c~\cite{Duplancic:2006nv}---together with possible soft-overlap and higher-Fock contributions beyond
the scope of the present leading-order twist-3 analysis.  In this respect
the pion electromagnetic form factors provide a useful benchmark within
the same framework.  Their NLO corrections have been computed in $k_{T}$
factorization up to the two-parton twist-3 level for both
spacelike~\cite{Cheng:2014gba} and timelike~\cite{Cheng:2015qra} momenta,
and the twist-2 and twist-3 NLO corrections partially cancel each other,
leaving a moderate net correction.  Two features of the present process,
however, caution against transferring this pattern directly.  The
two-meson final state involves two hard convolutions rather than one, and
the collinear NLO analysis of $\gam\gam\to M^{+}M^{-}$ indeed finds much
larger corrections than in the form-factor case.  Furthermore, in the
charge-suppressed neutral channels the twist-3 tower dominates the
amplitude, so the compensation between the twist-2 and twist-3 NLO terms
cannot operate in the same way, and the net correction is expected to be
governed by the twist-3 NLO correction alone.  An explicit NLO
$k_{T}$-factorization computation of these two-photon processes, while
beyond the scope of the present work, is therefore the natural next step
toward a quantitative description of the absolute cross sections.

\section{Summary}
\label{sec:summary}

We have computed the two-photon production of the neutral pseudoscalar
pairs $\gam\gam\to\pi^{0}\pi^{0}$ and $\gam\gam\to K_{S}^{0}K_{S}^{0}$ in
the $k_{T}$-factorization framework, including for the first time the
chirally enhanced two-parton twist-3 distribution amplitudes for these
channels.  The calculation sets up the LCDAs of $\pi^{0}$, $K^{0}$ and
$\bar K^{0}$, the twist-3 spinor projectors and light cone wave functions,
and the helicity hard kernels of the four diagrammatic groups, with the
$\KsKs$ channel obtained from $\KzKbz$ by the appropriate $CP$ projection.

The central physical message is that for these charge-suppressed channels
the subleading dynamics is not a correction but the leading effect.  The
cancellation of the $(e_{1}-e_{2})^{4}$ term removes the dominant
leading-twist amplitude, so that the cross sections are carried by the
intrinsic transverse-momentum and chirally enhanced twist-3 contributions.
Numerically the twist-3 cross section exceeds the twist-2 one throughout
the Belle window and brings the prediction much closer to the data.  The
calculation reproduces the angular shapes and, through the effective
power-law index $n_{\pi^{0}\pi^{0}}^{\rm tw3}=7.25$, the gross energy
dependence of the pion channel.  It does not, however, reproduce the very
steep fall of the neutral-kaon cross section, for which we obtain
$n_{K_{S}^{0}K_{S}^{0}}^{\rm tw3}=6.66$ against the measured
$n\simeq 10$--$11$, nor the pronounced
energy dependence of the measured cross section ratios.

These ratio observables, in which the leading dynamics largely cancels,
turn out to be the most discriminating.  Their strong, channel-dependent
energy variation points to the coexistence of competing mechanisms in the
few-GeV region, and we have shown that adding the soft handbag amplitude
coherently to the hard one---with a common form factor modulus and a
phase that rises with energy by roughly $50$--$100^{\circ}$ per GeV---reproduces the decrease of the pion ratio toward its minimum and the steep
fall of the kaon ratio.  In this picture the soft and hard amplitudes are of
comparable size in the opposite-helicity channel, so that both contribute
appreciably and neither can be neglected.  This interplay of hard and soft
dynamics is the central qualitative feature of the intermediate-energy
regime.

Two issues remain open.  First, even with the soft contribution the
absolute cross sections stay about a factor of three below the data, a
residual gap plausibly attributable to higher-order QCD corrections---the known NLO $K$-factor being sizable---supplemented possibly by
soft-overlap and higher-Fock contributions.  Second, the soft handbag
form factor is here fitted to the ratio data rather than derived, so its
modulus and phase await a first-principles understanding.  Progress on
both fronts---a next-to-leading-order $k_{T}$-factorization treatment of
the absolute normalization and a dynamical calculation of the soft form
factor---together with more precise data on the neutral channels at the
upper end of the Belle range and at Belle~II, would sharpen the separation
of the hard and soft components and provide a stringent test of the
picture advanced here.

\begin{acknowledgments}
This work was supported by the National Natural Science Foundation of
China under Grant No.~12305086, the Open Fund of the Key Laboratory of
Quark and Lepton Physics (MOE) under Grant No.~QLPL2024P01, and the
Project of Science and Technology Research Program of the Hubei
Provincial Department of Education under Grant No.~Q20222504.
\end{acknowledgments}

\section*{Data Availability}
There are no publicly available research data or software supporting this
manuscript.  Requests for further information or data should be sent to the
authors.

\bibliography{refs}

\end{document}